\documentclass{article}
\usepackage{iclr2027_conference,times}
\usepackage{graphicx}
\usepackage{booktabs}
\usepackage{multirow}
\usepackage{float}
\usepackage{caption}
\usepackage{wrapfig}
\usepackage{amsmath,amssymb}
\usepackage[hyperfootnotes=false]{hyperref}
\usepackage{url}
\usepackage{cleveref}
\usepackage{xspace}

\newcommand{\eg}{\emph{e.g.}\xspace}
\newcommand{\std}[1]{{\scriptsize #1}}
\newcommand{\suff}[1]{{\tiny #1}}

\hypersetup{
  hidelinks,
  pdftitle={UEPS: Robust and Efficient MRI Reconstruction},
  pdfauthor={Xiang Zhou, Hong Shang, Zijian Zhan, Tianyu He, Jintao Meng, Dong Liang}
}
\title{UEPS: Robust and Efficient MRI Reconstruction}
\author{
\bfseries Xiang Zhou$^{1,2,3,\star}$ \quad Hong Shang$^{2,\star}$ \quad Zijian Zhan$^{1,2,3}$ \\
\bfseries Tianyu He$^{2,4}$ \quad Jintao Meng$^{2,5,\dagger}$ \quad Dong Liang$^{2,6,\dagger}$ \\[0.6em]
\parbox{0.96\textwidth}{\normalfont\small\raggedright
$^{1}$School of Biomedical Engineering, Shenzhen University Medical School, Shenzhen University, China\\
$^{2}$Shenzhen Institutes of Advanced Technology, Chinese Academy of Sciences, China\\
$^{3}$Shenzhen University of Advanced Technology, China\\
$^{4}$University of Chinese Academy of Sciences, China\\
$^{5}$Faculty of Computer Science and Artificial Intelligence, Shenzhen University of Advanced Technology, China\\
$^{6}$State Key Laboratory of Biomedical Imaging Science and System, Chinese Academy of Sciences, China}
}

\iclrfinalcopy

\begin{document}
\maketitle
\lhead{arXiv preprint}
{\let\thefootnote\relax\footnotetext{$^{\star}$Equal contribution.}}
{\let\thefootnote\relax\footnotetext{$^{\dagger}$Corresponding authors: \texttt{\{jt.meng,dong.liang\}@siat.ac.cn}.}}

\begin{abstract}
  Deep unrolled models (DUM) have become the state-of-the-art for accelerated MRI reconstruction, yet their robustness under domain shift remains a critical barrier to clinical adoption. In this work, we identify coil sensitivity map (CSM) estimation as the primary bottleneck limiting generalization. To address this, we propose UEPS, a novel DUM architecture featuring three key designs: (i) an Unrolled Expanded (UE) design that eliminates CSM dependency by reconstructing each coil independently; (ii) progressive resolution, which leverages k-space-to-image mapping for efficient coarse-to-fine refinement; and (iii) sparse attention tailored to MRI's 1D undersampling nature. These physics-grounded designs enable gains in robustness and computational efficiency. We construct a large-scale zero-shot transfer benchmark comprising 10 out-of-distribution (OOD) test sets spanning diverse clinical shifts—anatomy, view, contrast, vendor, field strength, and coil configurations. Extensive experiments demonstrate that UEPS consistently and substantially outperforms existing DUM, end-to-end, diffusion, and untrained methods across all OOD tests, achieving state-of-the-art robustness with low-latency inference suitable for real-time deployment. Our code is available at \url{https://github.com/HongShangGroup/UEPS}.

\end{abstract}

\section{Introduction}

\suppressfloats[t]
\begin{figure}[t]
  \centering
  \includegraphics[height=4.8cm]{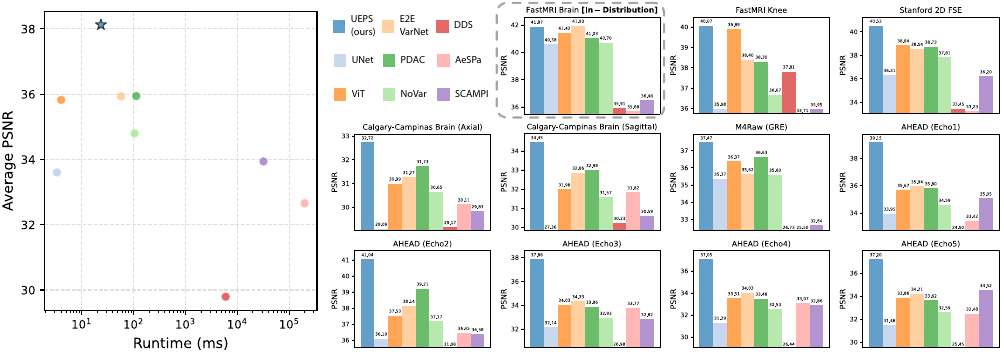}
  \caption{Zero-shot transfer benchmark results. Average PSNR: mean over 10 OOD sets. Runtime: measured on FastMRI Knee (15 coils). Compared methods include E2EVarNet (MICCAI 2020), PDAC (CVPR 2024), NoVar (CVPR 2025), DDS (ICLR 2024), AeSPa (CVPR 2025), and SCAMPI (MRM 2024), alongside UNet and ViT baselines.}
  \label{fig:benchmark_psnr}
\end{figure}

Magnetic resonance imaging (MRI) provides excellent soft-tissue contrast without ionizing radiation, but suffers from long acquisition times. Undersampled MRI aims to reconstruct high-fidelity images from a fraction of measurements in k-space to accelerate MRI scans. State-of-the-art methods predominantly use deep unrolled models (DUM) which integrate a physical forward model of the imaging process into learnable neural networks and are typically trained on large-scale datasets comprising multi-coil k-space measurements \citep{VarNet,E2EVarNet}.

Despite their success on benchmarks with abundant training data, DUM struggle with domain shift and transfer poorly to unseen acquisition protocols \mbox{\citep{Assessment,EvalRobust}}. To enhance robustness, recent studies have explored training DUM on combined datasets aggregated from multiple sources \citep{DiverseTrain}. However, MRI's wide contrast variability makes it difficult to curate a comprehensive training dataset. Untrained methods offer an alternative by avoiding dependence on training data and optimizing independently for each test case \citep{ZS-SSL,ConvDecoder,IMJENSE}. These methods are less susceptible to spurious correlations tied to a specific training distribution, but the limited information available in each undersampled input can constrain reconstruction quality \citep{MeasureRobust}. 

To address this robustness issue, we first identify the primary generalization bottleneck stems from the dependency of DUM on coil sensitivity maps (CSM). In multi-coil MRI, now the clinical standard, multiple receiver coils simultaneously acquire data, each with a characteristic CSM describing its spatial reception profile. CSM are used in DUM to convert individual coil images into a single image or vice versa. Our analysis reveals that learned CSM estimation becomes less reliable under domain shift, and the resulting inaccuracies can propagate through the CSM-dependent reconstruction pathway, potentially exacerbating reconstruction errors. We propose a simple yet effective redesign of DUM, termed Unrolled Expanded model (UE), which reconstructs each individual coil image separately, thereby eliminating the need for explicit CSM estimation. Reconstructing coil images before their final combination has precedents such as GrappaNet \citep{GrappaNet}; we are the first to systematically investigate independent coil-wise reconstruction for robustness under domain shift. UE gives up explicit modeling of inter-coil correlations, yet achieves comparable in-distribution (ID) performance and improved out-of-distribution (OOD) performance. However, processing coils independently increases computation during both training and inference. To make this design more scalable, we introduce two features for efficiency as follows.

Unlike typical image-to-image tasks like image restoration \citep{Swinir}, MRI reconstruction is a k-space-to-image mapping, suggesting that the standard down-sampling up-sampling architecture may be suboptimal here. We propose progressive resolution, whose core innovation lies in performing up-sampling by k-space expansion. Crucially, additional measurements are incorporated for the expanded k-space region, adding authentic high-frequency information, which cannot be achieved by image-space interpolation. In this up-sampling-only architecture, early stages run at lower resolution with substantially reduced computation. An additional benefit is the implicit curriculum learning: initial stages with denser k-space sampling pose an easier task, stabilizing training and providing improved initialization for subsequent high-resolution refinement. Sparse attention serves as another key feature for accelerating our transformer-based model. Recognizing that 1D MRI undersampling produces aliasing along a single axis, our sparse attention allocates more computation to this axis, achieving computational savings where full attention is unnecessary. Combining UE with progressive resolution and sparse attention yields UEPS.

In this work, we revisit the architecture of DUM to improve robustness under domain shifts in anatomy, view, contrast, vendor, field strength, and coil configuration while reducing reconstruction time. Our main contributions are as follows:

\begin{enumerate}
    \item Motivated by our analysis of CSM dependence as a generalization bottleneck, we propose UE that improves OOD reconstruction while maintaining comparable ID performance, albeit at increased computational cost.
    \item We further introduce progressive resolution and sparse attention, two design features grounded in the k-space-to-image mapping and 1D acceleration nature of MRI physics, which yield marked computational acceleration without compromising, and in fact improving, reconstruction quality.
    \item To rigorously assess generalization, we establish a large-scale zero-shot transfer benchmark and show that UEPS outperforms all competing approaches by a significant margin, while maintaining acceptable runtime.
\end{enumerate}

\section{Related Work}

\subsection{Deep Unrolled Models}
DUM represent a class of methods that unroll iterative optimization algorithms into network layers, combining the rigor of physics-based models with the representational power of deep learning \citep{VarNet,DeepCascade,MoDL}. The evolution of DUM methods has seen a shift from using pre-computed CSM from external tools \citep{ESPIRiT,bart} to jointly learning them with the reconstruction network \citep{E2EVarNet,Joint-ICNet}, a paradigm that underpins all contemporary state-of-the-art (SOTA) DUM methods. Subsequent research has explored diverse architectural innovations to further improve performance, such as convolutional recurrent network \citep{RecurrentVarNet}, hybrid convolutional-transformer network \citep{HUMUS-Net}, memory efficient transformer \citep{IRFRestormer}, neural operator-based architecture \citep{NoVar}, and leveraging adjacent slices input \citep{PromptMR+}. Despite architectural advances, these works predominantly report performance on ID tests of large datasets like FastMRI \citep{fastMRI}; the real-world robustness of these models remains an open question.

\subsection{Multi-Resolution}
Multi-resolution architectures are commonly used in low-level vision tasks such as image restoration and medical image reconstruction, leveraging coarse resolutions for computational efficiency and global context while preserving high-resolution pathways for spatial precision. In image restoration, early seminal works \citep{Uformer,Restormer} employ a transformer-based U-Shaped structure with downsampling and upsampling, to capture hierarchical multi-scale representations and maintain computational efficiency, which becomes gold standard for subsequent approaches \citep{AST-v2,SADT}. In DUM-based MRI reconstruction, each cascade contains a denoiser, typically with a multi-resolution architecture, ranging from UNet in early works \citep{E2EVarNet} to transformer-based \citep{IRFRestormer,SDUM} or hybrid convolution-transformer designs \citep{HUMUS-Net,PDAC} in recent approaches. Note this multi-resolution design is within each cascade. To the best of our knowledge, no prior work has explored multi-resolution across cascades. Learned upsampling operations are used in these multi-scale models, such as transposed convolutions \citep{TransConv} or pixel shuffle \citep{SubPixel}, which are well-suited for image-to-image tasks by learning to synthesize details. However, they cannot utilize the available high-frequency information present in k-space for MRI reconstruction. 

\subsection{Sparse Attention}
Sparse attention has become essential for scaling transformers to high-resolution images, where standard self-attention's quadratic complexity becomes prohibitive. Various strategies have been proposed to restrict attention to relevant subsets to reduce computational cost while largely preserving representational power. Early breakthroughs focused on fixed sparsity patterns like shifted non-overlapping local windows \citep{Swin}, sliding window \citep{NAT}, and factorized attention along axes \citep{Axial,Cswin}. Beyond static patterns, dynamic sparse attention predicts the most relevant tokens for each query based on the input content. For instance, DynamicViT progressively prunes redundant visual tokens using a lightweight prediction module \citep{Dynamicvit}, and SpargeAttn predicts the sparse mask by compressing each block of queries or keys to a single token \citep{Spargeattention}. Although sparse attention has been extensively studied for vision and language tasks, it remains unexplored in MRI reconstruction.

\begin{figure}[t]
  \centering
  \includegraphics[width=\linewidth]{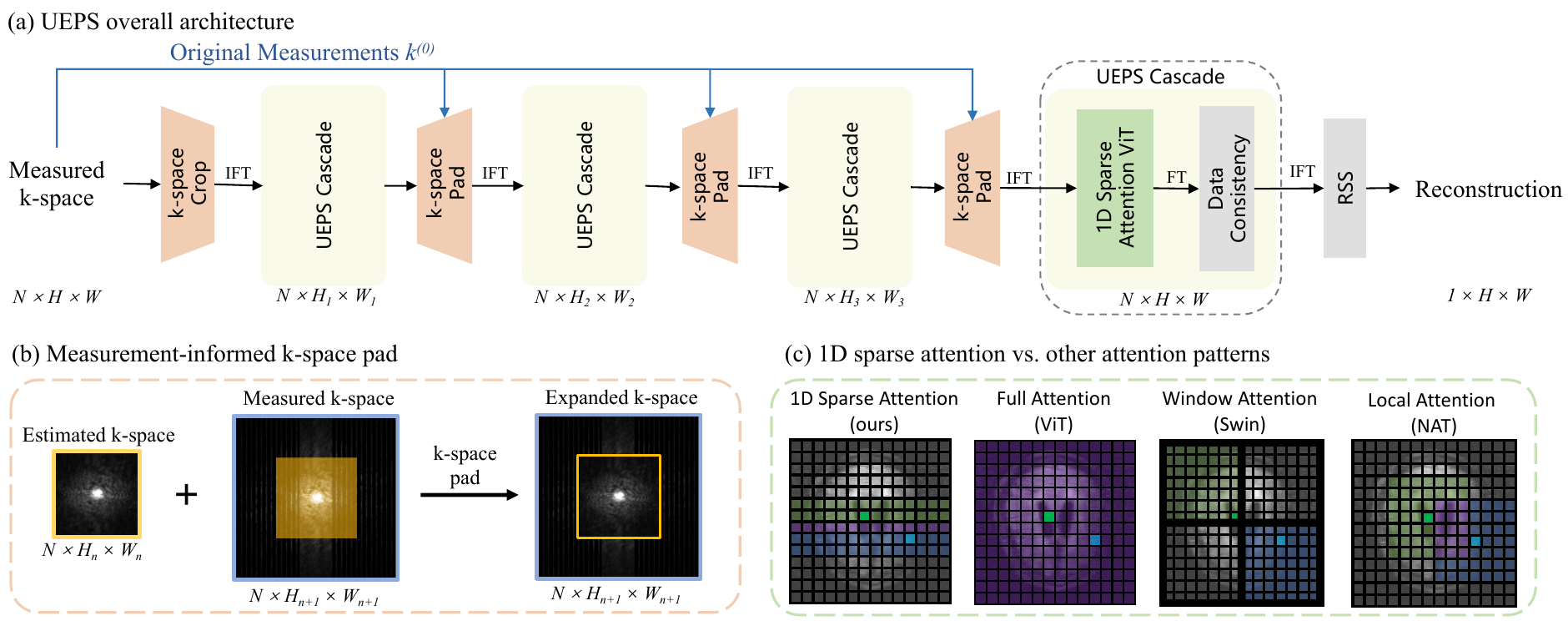}
  \caption{Overview of the proposed UEPS architecture, illustrating progressive resolution across cascades (e.g., 64 × 64, 128 × 128, 256 × 256 and 320 × 320) (a); schematic of measurement-informed k-space padding (b); and comparison of attention patterns, with different colours indicating the attention regions of different query tokens (c).}
  \label{fig:model}
\end{figure}

\section{Proposed Method}
UEPS, our proposed architecture, is a DUM redesigned for robustness without sacrificing efficiency. It consists of three main features,  the Unrolled Expanded model (UE), progressive resolution, and sparse attention, with the entire model shown in \cref{fig:model}. UE primarily improves robustness, but sacrifices speed. Progressive resolution and sparse attention, each motivated by MRI-specific principles, both reduce computation and further improve reconstruction quality.

\subsection{Problem Formulation}
\label{sec:formulation}
Modern MRI uses multiple receiver coils, each of which measures Fourier components of the imaged volume multiplied by a complex-valued position-dependent CSM \citep{fastMRI}. The measured k-space data from the $i$th coil, $\mathbf{k}_i$, are given by
\begin{equation}
  \mathbf{k}_i = \mathcal{F}(S_i\mathbf{m}) + \epsilon_i, \quad i = 1,2,\ldots,N
  \label{eq:fullsamp}
\end{equation}
where $\mathbf{m}$ is the imaging object, $S_i$ is the CSM of the $i$th coil, $\mathcal{F}$ is the Fourier transform operator, $\epsilon_i$ is measurement noise for the $i$th coil, and $N$ is the number of coils. For fully sampled k-space data, the $i$th coil image $\hat{\mathbf{m}}_i$ can be reconstructed by an inverse Fourier transform:
\begin{equation}
  \hat{\mathbf{m}}_i = \mathcal{F}^{-1}(\mathbf{k}_i)
  \label{eq:ift}
\end{equation}

Then, the individual coil images are combined with pixel-level root-sum-of-squares (RSS) to provide the final image estimate $\hat{\mathbf{m}}$ \citep{fastMRI}.
\begin{equation}
  \hat{\mathbf{m}} = \mathrm{RSS}(\hat{\mathbf{m}}_1, \dots, \hat{\mathbf{m}}_N) = \sqrt{\sum_{i = 1}^{N} |\hat{\mathbf{m}}_i|^2}
  \label{eq:rss}
\end{equation}

In undersampled MRI, only part of the k-space is acquired to accelerate scan
\begin{equation}
  \tilde{\mathbf{k}}_i = \mathcal{M}\mathcal{F}(S_i\mathbf{m}) + \epsilon_i, \quad i = 1,2,\ldots,N
  \label{eq:forwardmodel}
\end{equation}
where $\mathcal{M}$ is the binary sampling mask and $\tilde{\mathbf{k}}_i$ denotes the undersampled k-space data. DUM solves this ill-posed inverse problem of reconstructing undersampled MRI by integrating the physical forward model \cref{eq:fullsamp} into a cascaded neural network inspired by iterative optimization algorithms. Let $\mathbf{k}^{(0)} \in \mathbb{C}^{N \times H \times W}$ denote the Cartesian k-space data from all coils with unsampled points filled with zeros. $H$/$W$ is the number of readout samples/phase encoding in k-space, equivalent to height/width in image-space for Cartesian MRI. Let $\mathbf{x}^{(t)} \in \mathbb{C}^{N \times H \times W}$ denote the intermediate multi-coil image estimate at the $t$-th cascade of the unrolled model, and $\mathbf{k}^{(t)} \in \mathbb{C}^{N \times H \times W}$ the corresponding intermediate k-space estimate. Each refinement stage can be formulated as below with image-space intermediate quantities
\begin{align}
& \tilde{\mathbf{x}}^{(t)} = \mathbf{x}^{(t)} + \mathcal{E}(\mathcal{D}_{\theta}^{(t)}(\mathcal{R}(\mathbf{x}^{(t)}))) \label{eq:un1} \\
& \mathbf{k}^{(t)} = \mathcal{F}(\tilde{\mathbf{x}}^{(t)}) \label{eq:un2} \\
& \mathbf{k}^{(t+1)} = \mathcal{DC}_{\eta}(\mathbf{k}^{(t)}, \mathbf{k}^{(0)}) \label{eq:un3} \\
& \mathbf{x}^{(t+1)} = \mathcal{F}^{-1}(\mathbf{k}^{(t+1)}) \label{eq:un4}
\end{align}
where $\mathcal{R}$ and $\mathcal{E}$ are reduction and expansion operators to convert multi-coil images into a single image or vice versa by leveraging CSM with details given by \citet{E2EVarNet}. $\mathcal{D}_{\theta}^{(t)}$ is a learnable image denoiser, such as UNet \citep{E2EVarNet}, Restormer \citep{IRFRestormer}, or ViT \citep{ViT} used in this work. $\mathcal{DC}_{\eta}$ is the data consistency module with learnable weight $\eta$. While originally inspired by the gradient descent formulation of compressed sensing MRI, this module can accommodate various implementations, provided that data fidelity with acquired measurements is maintained. Notably, with the specific instantiation below with scalar $\eta$, \cref{eq:un1,eq:un2,eq:un3,eq:un4} are mathematically equivalent to the formulation of \citet{E2EVarNet}.
\begin{equation}
  \mathbf{k}^{(t+1)} = \mathcal{DC}_{\eta}(\mathbf{k}^{(t)}, \mathbf{k}^{(0)}) 
                     = \mathbf{k}^{(t)} - \eta\mathcal{M}(\mathbf{k}^{(t)} - \mathbf{k}^{(0)})
  \label{eq:dcsoft}
\end{equation}

$\mathbf{x}^{(0)}$, derived by \cref{eq:ift}, serves as an initial estimate with aliasing artifacts. The iterative updates concluded with coil combination step as in \cref{eq:rss}, form the complete MRI reconstruction pipeline.

\subsection{Unrolled Expanded Model}
UE reformulates reconstruction in terms of individual coil images, removing the CSM-based reduction and expansion operations in \cref{eq:un1}.
\begin{align}
& \tilde{\mathbf{x}}^{(t)} = \mathbf{x}^{(t)} + \mathcal{D}_{\theta}^{(t)}(\mathbf{x}^{(t)}) \\
& \mathbf{k}^{(t)} = \mathcal{F}(\tilde{\mathbf{x}}^{(t)}) \\
& \mathbf{k}^{(t+1)} = \mathcal{DC}_{\eta}(\mathbf{k}^{(t)}, \mathbf{k}^{(0)}) \\
& \mathbf{x}^{(t+1)} = \mathcal{F}^{-1}(\mathbf{k}^{(t+1)})
\end{align}
where $\mathcal{D}_{\theta}^{(t)}$ takes $\mathbf{x}^{(t)} \in \mathbb{C}^{N \times H \times W}$ as input, rather than $\mathcal{R}(\mathbf{x}^{(t)}) \in \mathbb{C}^{H \times W}$ in \cref{sec:formulation}. We treat multi coils as a batch dimension, keeping $\mathcal{D}_{\theta}^{(t)}$ unchanged for fair comparison with standard DUM. This design trades explicit modeling of inter-coil correlations for independence from estimated CSMs. This additional batch dimension increases computation, which is alleviated by the features introduced below.

\subsection{Progressive Resolution}
Progressive resolution is designed for efficient reconstruction of high-resolution images by delaying high-resolution processing to later stages, minimizing computation in early cascades. Our core innovation is performing up-sampling via k-space expansion, which populates the extended frequency region with actual high-frequency measurements that image-space methods cannot provide. This novel up-sampling operation integrates naturally into the unrolled pipeline, leading to the following formulation for each stage
\begin{align}
& \tilde{\mathbf{x}}^{(t)} = \mathbf{x}^{(t)} + \mathcal{D}_{\theta}^{(t)}(\mathbf{x}^{(t)}) \\
& \mathbf{k}^{(t)} = \mathcal{F}(\tilde{\mathbf{x}}^{(t)}) \\
& \tilde{\mathbf{k}}^{(t)} = \mathcal{DC}_{\eta}(\mathbf{k}^{(t)}, \mathbf{k}^{(0)}) \\
& \mathbf{k}^{(t+1)} = \mathcal{P}(\tilde{\mathbf{k}}^{(t)}, \mathbf{k}^{(0)}) \\
& \mathbf{x}^{(t+1)} = \mathcal{F}^{-1}(\mathbf{k}^{(t+1)})
\end{align}
where $\mathbf{x}^{(t)}, \tilde{\mathbf{x}}^{(t)}, \mathbf{k}^{(t)}, \tilde{\mathbf{k}}^{(t)} \in \mathbb{C}^{N \times h^{(t)} \times w^{(t)}}$, $h^{(t)}$ and $w^{(t)}$ define the progressive resolution schedule for the $t$-th stage. $\mathcal{P}$ is the k-space padding operator, filling the expanded region with acquired measurements where available, and zero elsewhere. Unlike UE and standard DUM, the initial estimate $\mathbf{x}^{(0)}$ is derived by \cref{eq:ift} after cropping $\mathbf{k}^{(0)}$ to the central region of size $h^{(0)} \times w^{(0)}$.

Almost all existing DUM approaches fall within a two-loop paradigm, an outer cascaded refinement loop with constant resolution, and an inner multi-resolution loop with a U-shaped backbone. Our progressive resolution design is, to the best of our knowledge, the first design to unify cascaded refinement and resolution scheduling into a single, up-sampling-only loop, which substantially improves efficiency by operating at lower resolutions for the bulk of the computation. Typical MRI sampling masks are denser in the center, leading to a higher sampling density of the initial low-resolution cascades with smaller k-space support, translating to a lower effective acceleration factor, making the early reconstruction considerably easier. Thus, our design inherently strengthens the progressive refinement nature of DUM.

\subsection{Sparse Attention}
Despite substantial acceleration from progressive resolution, the high-resolution cascades are still computationally heavy for high-resolution images, when employing a standard ViT as $\mathcal{D}_{\theta}$, whose full self-attention incurs quadratic complexity. Sparse attention is introduced by designing a sparsity pattern that determines which spatial locations attend to each other, computing attention only where it is meaningful and omitting it elsewhere to overcome the quadratic complexity.

We propose a novel sparsity pattern tailored for MRI, exploiting the fact that MRI undersampling, and thus aliasing, is anisotropic, a critical distinction from natural image tasks that assume isotropic spatial correlations. Acceleration in MRI is accomplished by undersampling along the phase-encode axis, while the readout direction is always fully sampled. From Fourier principles, multiplying k-space by a separable 1D mask corresponds to convolving the image with a 1D kernel, producing aliasing exclusively along that dimension. Following the convention of \citet{fastMRI}, we consider a 1D sampling mask along the $k_x$ axis, with the resulting aliasing along the $x$ axis.

\begin{wrapfigure}{r}{0.40\linewidth}
  \vspace{-1.2em}
  \centering
  \includegraphics[width=\linewidth]{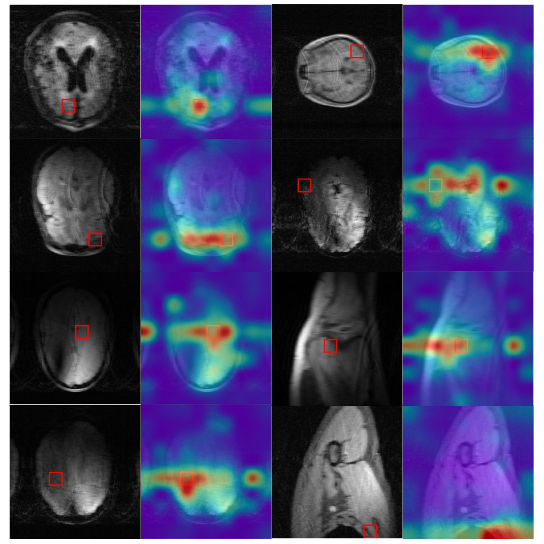}
  \caption{Attention maps from UE with a full-attention ViT backbone.}
  \label{fig:model_evidence}
  \vspace{-0.6em}
\end{wrapfigure}

Since the primary goal of $\mathcal{D}_{\theta}$ is to remove x-axis aliasing, meaningful attention is expected to occur predominantly along the x-axis. As shown in \cref{fig:model_evidence}, attention maps from UE with a standard ViT confirm this assumption, with scores concentrated along the x-axis. Rather than learning the sparsity pattern from data with a standard ViT, we design it explicitly to maximize efficiency. Concretely, we define that each token attends to all other tokens within the same row and its $n$ adjacent rows on each side, as shown in \cref{fig:model}. Additionally, we interleave sparse attention layers with full attention layers to preserve global context, following SOTA sparse attention designs in large language models \citep{Gemma2,Moba}. By faithfully leveraging MRI physical principles, namely k-space acquisition and 1D undersampling, UEPS achieves simultaneous gains in efficiency and performance, circumventing the typical trade-off between the two.

\section{Experiments and Results}

\subsection{Experimental Setups}
We construct a zero-shot transfer benchmark with ten OOD test sets from five public resources containing fully sampled multi-coil Cartesian k-space data (\cref{tab:dataset}). FastMRI Brain \citep{fastMRI}, the largest dataset among these resources, provides a data-rich training source for studying cross-dataset transfer. Its held-out split serves as the ID test set, while the remaining datasets evaluate OOD transfer across anatomy, view, sequence, contrast, vendor, field strength, and coil configuration without target-domain fine-tuning. All inputs are standardized to $320\times320$ through image-space cropping or padding, following \citet{HUMUS-Net,PDAC}.

The benchmark includes DUM \citep{E2EVarNet,PDAC,NoVar}, end-to-end networks \citep{Unet,MRViT}, diffusion-based reconstruction \citep{DDS}, and untrained methods \citep{SCAMPI,Aespa}. DUM and end-to-end baselines are retrained to match our settings; DDS uses released pretrained weights, and untrained methods are directly applied. Benchmark baselines retain their original architectural configurations, including cascade counts where applicable, while UEPS uses four cascades. Controlled ablations use matched configurations and training protocols, as detailed in \cref{sec:additional_setups}.

Unless otherwise specified, we retrospectively undersample k-space using equispaced masks with $4\times$ acceleration and an 8\% fully sampled center \citep{fastMRI}. We fix the acceleration factor to focus on cross-dataset transfer; substantial OOD degradation is already observed at $4\times$. We evaluate PSNR and SSIM against fully sampled RSS references, using the reference volume maximum for normalization. Runtime is measured on a single RTX 5090 GPU unless noted. DDS follows its original evaluation protocol, using Gaussian sampling masks and an MVUE reference \citep{DDS}.

\subsection{Evaluation of CSM Estimation}
\label{sec:csm_generalization}

\begin{figure}[tb]
  \centering

  \begin{minipage}{\linewidth}
    \centering
    \includegraphics[width=0.6\linewidth]{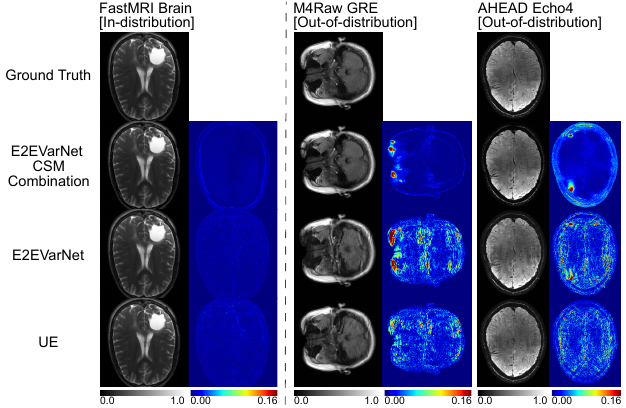}
    \caption{CSM combination and reconstruction on ID and OOD examples. CSM combination uses E2EVarNet-estimated CSMs to combine fully sampled coil images; E2EVarNet and UE reconstructions use undersampled inputs. All error maps are computed relative to the RSS reference.}
    \label{fig:csm}
  \end{minipage}

  \vspace{0.5em}
  \begin{minipage}[t]{0.56\linewidth}
    \vspace{0pt}
    \centering
    \captionsetup{type=table,skip=2pt}
    \caption{CSM combination comparison.}
    \label{tab:rss_mvue}
    \scriptsize
    \setlength{\tabcolsep}{3pt}
    \renewcommand{\arraystretch}{0.95}
    \resizebox{\linewidth}{!}{%
      \begin{tabular}{l l c c c c}
        \toprule
        Dataset & Metric & BART & E2EVarNet & PDAC & NoVar \\
        \midrule
        \multirow{2}{*}{\parbox{1.5cm}{FastMRI Brain [ID]}}
        & PSNR$\uparrow$ & \textbf{44.95} & \underline{44.28} & 43.84 & 41.98 \\
        & SSIM$\uparrow$ & \textbf{99.00} & \underline{98.97} & 98.85 & 98.85 \\
        \midrule
        \multirow{2}{*}{\parbox{1.5cm}{Average [OOD]}}
        & PSNR$\uparrow$ & \textbf{41.71} & \underline{38.16} & 36.82 & 37.09 \\
        & SSIM$\uparrow$ & \textbf{96.50} & \underline{94.44} & 92.24 & 94.31 \\
        \bottomrule
      \end{tabular}%
    }
  \end{minipage}\hfill
  \begin{minipage}[t]{0.41\linewidth}
    \vspace{0pt}
    \centering
    \captionsetup{type=table,skip=2pt}
    \caption{Reconstruction comparison.}
    \label{tab:e2evarnet_ueunet}
    \scriptsize
    \setlength{\tabcolsep}{3pt}
    \renewcommand{\arraystretch}{0.95}
    \resizebox{\linewidth}{!}{%
      \begin{tabular}{l l c c}
        \toprule
        Dataset & Metric & E2EVarNet & UE \\
        \midrule
        \multirow{2}{*}{\parbox{1.5cm}{FastMRI Brain [ID]}}
        & PSNR$\uparrow$ & \textbf{41.93} & 41.90 \\
        & SSIM$\uparrow$ & \textbf{96.27} & 96.16 \\
        \midrule
        \multirow{2}{*}{\parbox{1.5cm}{Average [OOD]}}
        & PSNR$\uparrow$ & 35.33 & \textbf{37.54} \\
        & SSIM$\uparrow$ & 90.70 & \textbf{92.82} \\
        \bottomrule
      \end{tabular}%
    }
  \end{minipage}
\end{figure}

As ground-truth CSM are unavailable, we assess CSM quality by comparing the CSM-combined multi-coil image (CSM combination) with the RSS combination, which an ideal CSM should recover. BART \citep{bart}  is a well-established untrained baseline for CSM estimation, providing a consistent reference for assessing the intrinsic difficulty of CSM estimation across datasets. As shown in \cref{tab:rss_mvue}, E2EVarNet performs comparably to BART on ID data, but their PSNR gap widens from 0.67 dB to 3.55 dB on average across OOD test sets. PDAC and NoVar also fall substantially behind BART on OOD data, highlighting the limited cross-dataset generalization of learned CSM estimation. \Cref{fig:csm} shows visual examples of how CSM errors relate to reconstruction errors. In the OOD examples, localized CSM estimation errors propagate through the reconstruction network, resulting in substantial reconstruction failure in the same regions.  UE, without explicit CSM estimation, avoids CSM-induced regional artifacts. This observation is supported by the quantitative comparison in \cref{tab:e2evarnet_ueunet}: UE maintains comparable ID performance while improving average OOD PSNR by 2.21 dB. Together, these results identify dependence on learned CSM estimation as a bottleneck to OOD generalization.

\subsection{Zero-shot Transfer Benchmark Results}
\begin{figure}[tb]
  \centering
  \includegraphics[height=7.5cm]{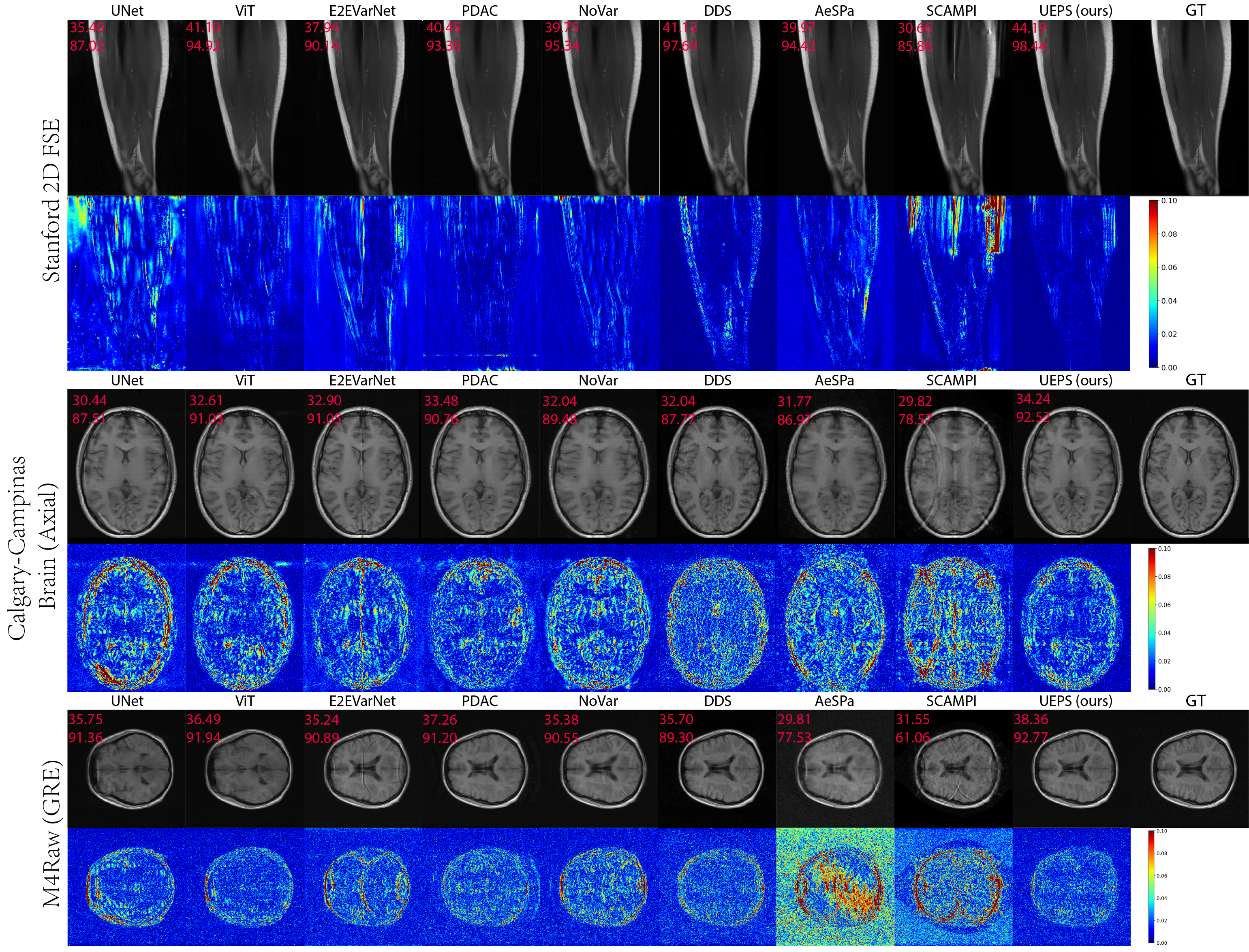}
  \caption{Reconstruction examples from the zero-shot transfer benchmark, each with PSNR/SSIM shown in the top-left corner.}
  \label{fig:examples}
\end{figure}

\Cref{fig:benchmark_psnr}, together with the qualitative comparison in \cref{fig:examples}, provides clear evidence that UEPS markedly improves robustness, consistently surpassing existing SOTA methods across all OOD test sets and achieving a large performance margin in aggregate, with complete results in \cref{tab:supp_full_benchmark}. The strong ID performance of other DUM methods and end-to-end methods fails to generalize to OOD settings, underscoring the need for zero-shot transfer evaluation for real-world deployment. Untrained methods trade off ID performance as expected; however, this sacrifice brings no consistent OOD advantage over supervised techniques. Remarkably, UEPS delivers robust performance with low latency, enabled by two efficiency-driven features that make real-time deployment feasible.

\subsection{Ablation Studies}

\begin{wrapfigure}{r}{0.46\linewidth}
  \centering
  \includegraphics[width=\linewidth]{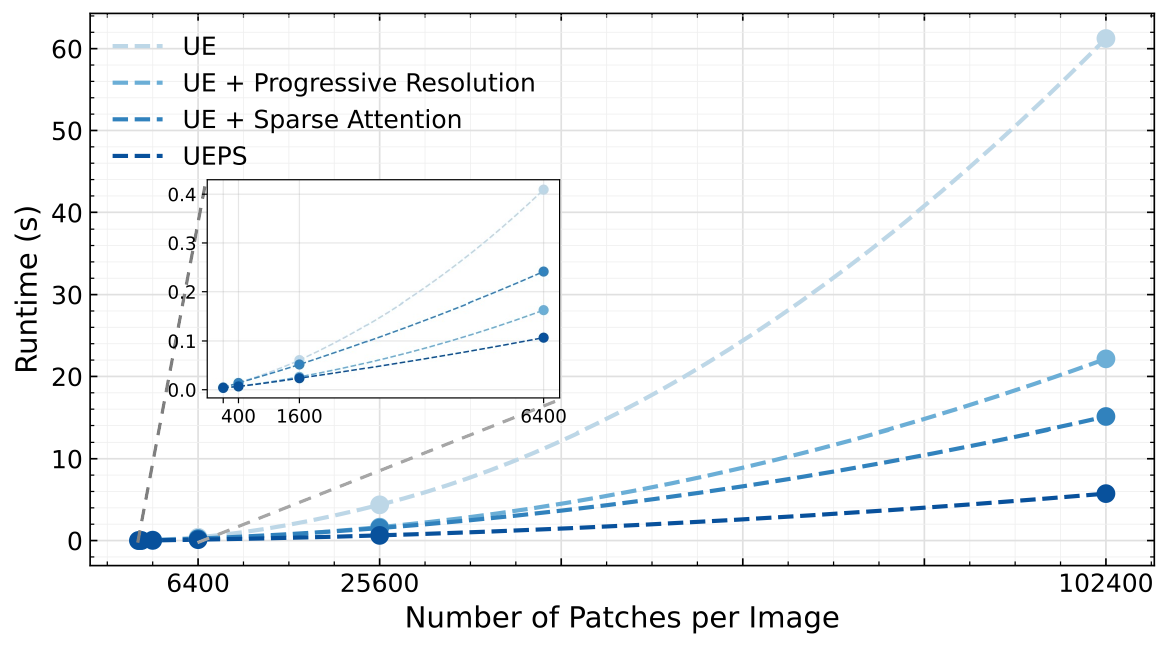}
  \caption{Runtime comparison.}
  \label{fig:speed}
\end{wrapfigure}
We evaluate progressive resolution and sparse attention using the same ViT-based UE framework, adding each independently, and combining both as UEPS. All four configurations use the same training protocol and comparable parameter budgets. As shown in \cref{tab:ablation}, both components improve average reconstruction quality and reduce latency relative to UE. Incorporating progressive resolution with sparse attention substantially reduces inference latency with comparable reconstruction quality. \Cref{fig:speed} illustrates their complementary benefits: progressive resolution reduces computation across cascades, while sparse attention limits attention cost as the token count increases.

\subsection{Multi-Resolution Comparison}

We evaluate our progressive resolution design against competitive multi-resolution architectures from both the general vision domain and prior MRI reconstruction studies. We select Restormer \citep{Restormer}, a cornerstone in image restoration with a U-shaped transformer, along with a recent improved variant \citep{AST-v2}. Given the high memory footprint of these two models, we use a reduced configuration with approximately 30M parameters and a single UE cascade, the maximum feasible within our GPU memory. Two prior multi-resolution MRI methods are included: GrappaNet \citep{GrappaNet}, a UNet-based coil-wise approach similar to our UE, and PDAC \citep{PDAC}, a DUM approach with inner-cascade multi-resolution. GrappaNet and our model are likewise configured with 30M parameters for fair comparison, while PDAC uses its original configuration with roughly 130M parameters.

\Cref{tab:uep} shows our model achieves top robustness, followed by Restormer and ASTv2, while PDAC lags due to its fragile CSM estimation. The strong representational capacity of Restormer and ASTv2 comes at the cost of substantial computational burden and GPU memory consumption. In contrast, UEPS  merges cascaded refinement with progressive resolution scheduling in a single, up-sampling-only loop, thus concentrating most computation at low resolutions for superior efficiency. Note that UEPS performance in \cref{tab:uep} is worse than that in \cref{tab:ablation} due to reduced model size.

\subsection{Sparse Attention Comparison}
We compare our design with static sparse attention methods, Swin \citep{Swin}, CSWin \citep{Cswin}, and NAT \citep{NAT}, and the dynamic method Sparge \citep{Spargeattention}. Representative patterns appear in \cref{fig:model}(c). All methods use UE with progressive resolution, varying only the attention mechanism. Note that the runtime of Sparge is measured on a 4090 GPU and converted to a 5090-equivalent estimate because of hardware incompatibility. As shown in \cref{tab:sparseattention}, our MRI-tailored sparse attention outperforms general vision counterparts by explicitly leveraging MRI's 1D aliasing nature.

\begin{table}[t!]
\centering
\caption{Ablation of progressive resolution and sparse attention. The final row corresponds to UEPS. Best values are bold and second best underlined.}
\label{tab:ablation}
\begin{minipage}{\linewidth}
  \centering
  \fontsize{7pt}{8.5pt}\selectfont
  \setlength{\tabcolsep}{0.6pt}
  \renewcommand{\arraystretch}{1.12}
  \begin{tabular*}{\linewidth}{@{\extracolsep{\fill}}c c c @{\hspace{2pt}}c c c @{\hspace{2pt}}c c c @{\hspace{2pt}}c c c @{\hspace{2pt}}c c c@{}}
  \toprule
  \multirow{2}{*}{UE} & \multirow{2}{*}{\shortstack{Progressive\\resolution}} & \multirow{2}{*}{\shortstack{Sparse\\attention}} & \multicolumn{3}{c}{M4Raw GRE} & \multicolumn{3}{c}{FastMRI Knee} & \multicolumn{3}{c}{AHEAD Echo3} & \multicolumn{3}{c}{Average} \\
  \cmidrule(lr){4-6}\cmidrule(lr){7-9}\cmidrule(lr){10-12}\cmidrule(l){13-15}
  & & & PSNR & SSIM & Runtime & PSNR & SSIM & Runtime & PSNR & SSIM & Runtime & PSNR & SSIM & Runtime \\
  \midrule
  $\checkmark$ & --- & --- & \underline{37.80} & 91.93 & 16.51 & 39.91 & 92.08 & 60.78 & 37.34 & 93.11 & 132.56 & 37.90 & 93.08 & 97.28 \\
  $\checkmark$ & $\checkmark$ & --- & 37.62 & \underline{92.05} & \underline{7.31} & \underline{39.97} & 92.31 & \underline{26.29} & 37.57 & \underline{93.41} & \underline{57.93} & 37.98 & 93.25 & \underline{42.32} \\
  \addlinespace[2pt]
  $\checkmark$ & --- & $\checkmark$ & \textbf{37.81} & \textbf{92.11} & 13.71 & \textbf{40.07} & \underline{92.37} & 51.77 & \underline{37.75} & \textbf{93.66} & 111.01 & \textbf{38.19} & \textbf{93.47} & 81.28 \\
  \addlinespace[2pt]
  $\checkmark$ & $\checkmark$ & $\checkmark$ & 37.47 & 91.94 & \textbf{6.50} & \textbf{40.07} & \textbf{92.38} & \textbf{23.61} & \textbf{37.86} & \textbf{93.66} & \textbf{51.85} & \underline{38.13} & \underline{93.39} & \textbf{38.02} \\
  \bottomrule
  \end{tabular*}
\end{minipage}

\vspace{0.8em}

\begin{minipage}[t]{0.515\linewidth}
  \vspace{0pt}
  \centering
  \caption{Multi-resolution comparison.}
  \label{tab:uep}
  \scriptsize
  \setlength{\tabcolsep}{2pt}
  \renewcommand{\arraystretch}{0.92}
  \resizebox{\linewidth}{!}{%
  \begin{tabular}{l l c c c c c}
  \toprule
   & & \multicolumn{5}{c}{Model} \\
  \cmidrule{3-7}
  Dataset & Metric & GrappaNet & Restormer & ASTv2 & PDAC & UEPS\suff{(Ours)} \\
  \midrule
  \multirow{3}{*}{\parbox{1.2cm}{M4Raw GRE}}
  & PSNR$\uparrow$ & 36.40 & 36.50 & \underline{36.93} & 36.63 & \textbf{37.32} \\
  & SSIM$\uparrow$ & 91.01 & 91.30 & \underline{91.59} & 90.31 & \textbf{91.76} \\
  & Runtime$\downarrow$ & \underline{4.96} & 107.34 & 87.21 & 98.99 & \textbf{2.32} \\
  \midrule
  \multirow{3}{*}{\parbox{1.2cm}{FastMRI Knee}}
  & PSNR$\uparrow$ & 38.88 & 39.83 & \textbf{39.90} & 38.30 & \underline{39.87} \\
  & SSIM$\uparrow$ & 91.54 & \underline{92.38} & \textbf{92.40} & 90.38 & 92.08 \\
  & Runtime$\downarrow$ & \underline{15.68} & 194.67 & 185.55 & 113.22 & \textbf{7.43} \\
  \midrule
  \multirow{3}{*}{\parbox{1.2cm}{AHEAD Echo3}}
  & PSNR$\uparrow$ & 35.34 & \underline{35.82} & 35.38 & 33.86 & \textbf{37.24} \\
  & SSIM$\uparrow$ & 91.41 & \underline{91.94} & 91.45 & 87.85 & \textbf{93.01} \\
  & Runtime$\downarrow$ & \underline{39.44} & 755.35 & 607.07 & 122.62 & \textbf{16.14} \\
  \midrule
  \multirow{3}{*}{Average}
  & PSNR$\uparrow$ & 36.03 & 36.71 & \underline{36.72} & 35.94 & \textbf{37.57} \\
  & SSIM$\uparrow$ & 91.59 & \underline{92.24} & 92.10 & 90.28 & \textbf{92.76} \\
  & Runtime$\downarrow$ & \underline{28.66} & 543.00 & 439.88 & 114.94 & \textbf{11.85} \\
  \bottomrule
  \end{tabular}%
  }
\end{minipage}\hfill
\begin{minipage}[t]{0.455\linewidth}
  \vspace{0pt}
  \centering
  \caption{Sparse-attention comparison.}
  \label{tab:sparseattention}
  \scriptsize
  \setlength{\tabcolsep}{2pt}
  \renewcommand{\arraystretch}{0.92}
  \resizebox{\linewidth}{!}{%
  \begin{tabular}{l l c c c c c}
  \toprule
   & & \multicolumn{5}{c}{Model} \\
  \cmidrule{3-7}
  Dataset & Metric & Swin & CSWin & NAT & Sparge & UEPS\suff{(Ours)} \\
  \midrule
  \multirow{3}{*}{\parbox{1.3cm}{M4Raw GRE}}
  & PSNR$\uparrow$ & 37.18 & \underline{37.47} & \textbf{37.59} & 36.58 & \underline{37.47} \\
  & SSIM$\uparrow$ & 91.61 & \textbf{91.95} & 91.92 & 91.09 & \underline{91.94} \\
  & Runtime$\downarrow$ & 6.91 & \underline{6.58} & 6.77 & 8.60 & \textbf{6.50} \\
  \midrule
  \multirow{3}{*}{\parbox{1.3cm}{FastMRI Knee}}
  & PSNR$\uparrow$ & 39.94 & 39.90 & \underline{39.98} & 39.71 & \textbf{40.07} \\
  & SSIM$\uparrow$ & 92.29 & 92.20 & \underline{92.35} & 92.10 & \textbf{92.38} \\
  & Runtime$\downarrow$ & 24.60 & \textbf{23.54} & 24.14 & 29.79 & \underline{23.61} \\
  \midrule
  \multirow{3}{*}{\parbox{1.3cm}{AHEAD Echo3}}
  & PSNR$\uparrow$ & 37.05 & \underline{37.60} & 37.25 & 36.98 & \textbf{37.86} \\
  & SSIM$\uparrow$ & 92.94 & \underline{93.43} & 93.00 & 92.69 & \textbf{93.66} \\
  & Runtime$\downarrow$ & 53.93 & \textbf{51.15} & 53.25 & 64.72 & \underline{51.85} \\
  \midrule
  \multirow{3}{*}{Average}
  & PSNR$\uparrow$ & 37.62 & \underline{38.01} & 37.80 & 37.36 & \textbf{38.13} \\
  & SSIM$\uparrow$ & 92.92 & \underline{93.28} & 93.07 & 92.56 & \textbf{93.39} \\
  & Runtime$\downarrow$ & 39.72 & \textbf{37.41} & 38.96 & 47.57 & \underline{38.02} \\
  \bottomrule
  \end{tabular}%
  }
\end{minipage}
\end{table}



\section{Limitations}
Our study focuses on architectural robustness, leaving diverse training data, which improve robustness in MRI \citep{DiverseTrain,DataFilter} and vision \citep{CLIP,SAM}, for future integration. Generalization to unseen sampling masks and acceleration factors remains untested. Prior work indicates this could be alleviated through training with diverse sampling patterns \citep{DiverseTrain}. We also focus on 2D reconstruction. Reported benefits of 3D inputs \citep{PromptMR+} motivate extending our sparse attention to volumetric reconstruction.

\section{Conclusion}
We significantly improve MRI reconstruction robustness through three MRI-grounded design features: unrolled expanded model, progressive resolution, and sparse attention. These innovations yield simultaneous gains in performance and speed. A large-scale zero-shot transfer benchmark across 10 diverse OOD datasets confirms that UEPS consistently and substantially outperforms existing methods, establishing a new standard for robust real-world MRI reconstruction.

\subsection*{AI use statement}
Generative AI was used to assist with language polishing and improve the clarity of the manuscript. All revisions were reviewed and verified by the authors.

\bibliographystyle{iclr2027_conference}
\bibliography{main}

\begin{thebibliography}{56}
\providecommand{\natexlab}[1]{#1}
\providecommand{\url}[1]{\texttt{#1}}
\expandafter\ifx\csname urlstyle\endcsname\relax
  \providecommand{\doi}[1]{doi: #1}\else
  \providecommand{\doi}{doi: \begingroup \urlstyle{rm}\Url}\fi

\bibitem[Aggarwal et~al.(2018)Aggarwal, Mani, and Jacob]{MoDL}
Hemant~K Aggarwal, Merry~P Mani, and Mathews Jacob.
\newblock Modl: Model-based deep learning architecture for inverse problems.
\newblock \emph{IEEE transactions on medical imaging}, 38\penalty0
  (2):\penalty0 394--405, 2018.

\bibitem[Alkemade et~al.(2020)Alkemade, Mulder, Groot, Isaacs, van Berendonk,
  Lute, Isherwood, Bazin, and Forstmann]{AHEAD}
Anneke Alkemade, Martijn~J Mulder, Josephine~M Groot, Bethany~R Isaacs, Nikita
  van Berendonk, Nicky Lute, Scott~JS Isherwood, Pierre-Louis Bazin, and
  Birte~U Forstmann.
\newblock The amsterdam ultra-high field adult lifespan database (ahead): A
  freely available multimodal 7 tesla submillimeter magnetic resonance imaging
  database.
\newblock \emph{NeuroImage}, 221:\penalty0 117200, 2020.

\bibitem[Cheng(2018)]{stanford2d}
Joseph Cheng.
\newblock Stanford 2d fse dataset.
\newblock mridata.org, 2018.
\newblock URL \url{http://mridata.org/list?project=Stanford%202D%20FSE}.

\bibitem[Chung et~al.(2023)Chung, Lee, and Ye]{DDS}
Hyungjin Chung, Suhyeon Lee, and Jong~Chul Ye.
\newblock Decomposed diffusion sampler for accelerating large-scale inverse
  problems.
\newblock \emph{arXiv preprint arXiv:2303.05754}, 2023.

\bibitem[Darestani \& Heckel(2021)Darestani and Heckel]{ConvDecoder}
Mohammad~Zalbagi Darestani and Reinhard Heckel.
\newblock Accelerated mri with un-trained neural networks.
\newblock \emph{IEEE Transactions on Computational Imaging}, 7:\penalty0
  724--733, 2021.

\bibitem[Darestani et~al.(2021)Darestani, Chaudhari, and Heckel]{MeasureRobust}
Mohammad~Zalbagi Darestani, Akshay~S Chaudhari, and Reinhard Heckel.
\newblock Measuring robustness in deep learning based compressive sensing.
\newblock In \emph{International Conference on Machine Learning}, pp.\
  2433--2444. PMLR, 2021.

\bibitem[Darestani et~al.(2024)Darestani, Nath, Li, He, Roth, Xu, Xu, Heckel,
  and Zhao]{IRFRestormer}
Mohammad~Zalbagi Darestani, Vishwesh Nath, Wenqi Li, Yufan He, Holger~R Roth,
  Ziyue Xu, Daguang Xu, Reinhard Heckel, and Can Zhao.
\newblock Ir-frestormer: Iterative refinement with fourier-based restormer for
  accelerated mri reconstruction.
\newblock In \emph{Proceedings of the IEEE/CVF winter conference on
  applications of computer vision}, pp.\  7655--7664, 2024.

\bibitem[Dong et~al.(2024)Dong, Feng, Guessous, Liang, and He]{FlexAttn}
Juechu Dong, Boyuan Feng, Driss Guessous, Yanbo Liang, and Horace He.
\newblock Flex attention: A programming model for generating optimized
  attention kernels.
\newblock \emph{arXiv preprint arXiv:2412.05496}, 2\penalty0 (3):\penalty0 4,
  2024.

\bibitem[Dong et~al.(2022)Dong, Bao, Chen, Zhang, Yu, Yuan, Chen, and
  Guo]{Cswin}
Xiaoyi Dong, Jianmin Bao, Dongdong Chen, Weiming Zhang, Nenghai Yu, Lu~Yuan,
  Dong Chen, and Baining Guo.
\newblock Cswin transformer: A general vision transformer backbone with
  cross-shaped windows.
\newblock In \emph{Proceedings of the IEEE/CVF conference on computer vision
  and pattern recognition}, pp.\  12124--12134, 2022.

\bibitem[Dosovitskiy et~al.(2020)Dosovitskiy, Beyer, Kolesnikov, Weissenborn,
  Zhai, Unterthiner, Dehghani, Minderer, Heigold, Gelly, et~al.]{ViT}
Alexey Dosovitskiy, Lucas Beyer, Alexander Kolesnikov, Dirk Weissenborn,
  Xiaohua Zhai, Thomas Unterthiner, Mostafa Dehghani, Matthias Minderer, Georg
  Heigold, Sylvain Gelly, et~al.
\newblock An image is worth 16x16 words: Transformers for image recognition at
  scale.
\newblock \emph{arXiv preprint arXiv:2010.11929}, 2020.

\bibitem[Dumoulin \& Visin(2016)Dumoulin and Visin]{TransConv}
Vincent Dumoulin and Francesco Visin.
\newblock A guide to convolution arithmetic for deep learning.
\newblock \emph{arXiv preprint arXiv:1603.07285}, 2016.

\bibitem[Fabian et~al.(2022)Fabian, Tinaz, and Soltanolkotabi]{HUMUS-Net}
Zalan Fabian, Berk Tinaz, and Mahdi Soltanolkotabi.
\newblock Humus-net: Hybrid unrolled multi-scale network architecture for
  accelerated mri reconstruction.
\newblock \emph{Advances in Neural Information Processing Systems},
  35:\penalty0 25306--25319, 2022.

\bibitem[Fang et~al.(2024)Fang, Sun, Wang, Huang, Wang, and Cao]{Eva02}
Yuxin Fang, Quan Sun, Xinggang Wang, Tiejun Huang, Xinlong Wang, and Yue Cao.
\newblock Eva-02: A visual representation for neon genesis.
\newblock \emph{Image and Vision Computing}, 149:\penalty0 105171, 2024.

\bibitem[Feng et~al.(2023)Feng, Wu, Feng, She, Liu, Zhang, and Wei]{IMJENSE}
Ruimin Feng, Qing Wu, Jie Feng, Huajun She, Chunlei Liu, Yuyao Zhang, and
  Hongjiang Wei.
\newblock Imjense: scan-specific implicit representation for joint coil
  sensitivity and image estimation in parallel mri.
\newblock \emph{IEEE Transactions on Medical Imaging}, 43\penalty0
  (4):\penalty0 1539--1553, 2023.

\bibitem[Hammernik et~al.(2018)Hammernik, Klatzer, Kobler, Recht, Sodickson,
  Pock, and Knoll]{VarNet}
Kerstin Hammernik, Teresa Klatzer, Erich Kobler, Michael~P Recht, Daniel~K
  Sodickson, Thomas Pock, and Florian Knoll.
\newblock Learning a variational network for reconstruction of accelerated mri
  data.
\newblock \emph{Magnetic resonance in medicine}, 79\penalty0 (6):\penalty0
  3055--3071, 2018.

\bibitem[Hassani et~al.(2023)Hassani, Walton, Li, Li, and Shi]{NAT}
Ali Hassani, Steven Walton, Jiachen Li, Shen Li, and Humphrey Shi.
\newblock Neighborhood attention transformer.
\newblock In \emph{Proceedings of the IEEE/CVF conference on computer vision
  and pattern recognition}, pp.\  6185--6194, 2023.

\bibitem[He et~al.(2025)He, Quan, Xu, and Ji]{SADT}
Xuyi He, Yuhui Quan, Ruotao Xu, and Hui Ji.
\newblock A universal scale-adaptive deformable transformer for image
  restoration across diverse artifacts.
\newblock In \emph{Proceedings of the IEEE/CVF Conference on Computer Vision
  and Pattern Recognition}, pp.\  12731--12741, 2025.

\bibitem[Jatyani et~al.(2025)Jatyani, Wang, Chandrashekar, Wu, Liu-Schiaffini,
  Tolooshams, and Anandkumar]{NoVar}
Armeet~Singh Jatyani, Jiayun Wang, Aditi Chandrashekar, Zihui Wu, Miguel
  Liu-Schiaffini, Bahareh Tolooshams, and Anima Anandkumar.
\newblock A unified model for compressed sensing mri across undersampling
  patterns.
\newblock In \emph{Proceedings of the Computer Vision and Pattern Recognition
  Conference}, pp.\  26004--26013, 2025.

\bibitem[Johnson et~al.(2021)Johnson, Jeong, Hammernik, Schlemper, Qin, Duan,
  Rueckert, Lee, Pezzotti, De~Weerdt, et~al.]{EvalRobust}
Patricia~M Johnson, Geunu Jeong, Kerstin Hammernik, Jo~Schlemper, Chen Qin,
  Jinming Duan, Daniel Rueckert, Jingu Lee, Nicola Pezzotti, Elwin De~Weerdt,
  et~al.
\newblock Evaluation of the robustness of learned mr image reconstruction to
  systematic deviations between training and test data for the models from the
  fastmri challenge.
\newblock In \emph{International Workshop on Machine Learning for Medical Image
  Reconstruction}, pp.\  25--34. Springer, 2021.

\bibitem[Joo et~al.(2025)Joo, Kim, Won, Lee, Eo, and Hwang]{Aespa}
Jinho Joo, Hyeseong Kim, Hyeyeon Won, Deukhee Lee, Taejoon Eo, and Dosik Hwang.
\newblock Aespa: Attention-guided self-supervised parallel imaging for mri
  reconstruction.
\newblock In \emph{Proceedings of the Computer Vision and Pattern Recognition
  Conference}, pp.\  5217--5226, 2025.

\bibitem[Jun et~al.(2021)Jun, Shin, Eo, and Hwang]{Joint-ICNet}
Yohan Jun, Hyungseob Shin, Taejoon Eo, and Dosik Hwang.
\newblock Joint deep model-based mr image and coil sensitivity reconstruction
  network (joint-icnet) for fast mri.
\newblock In \emph{Proceedings of the IEEE/CVF conference on computer vision
  and pattern recognition}, pp.\  5270--5279, 2021.

\bibitem[Kirillov et~al.(2023)Kirillov, Mintun, Ravi, Mao, Rolland, Gustafson,
  Xiao, Whitehead, Berg, Lo, et~al.]{SAM}
Alexander Kirillov, Eric Mintun, Nikhila Ravi, Hanzi Mao, Chloe Rolland, Laura
  Gustafson, Tete Xiao, Spencer Whitehead, Alexander~C Berg, Wan-Yen Lo, et~al.
\newblock Segment anything.
\newblock In \emph{Proceedings of the IEEE/CVF international conference on
  computer vision}, pp.\  4015--4026, 2023.

\bibitem[Knoll et~al.(2019)Knoll, Hammernik, Kobler, Pock, Recht, and
  Sodickson]{Assessment}
Florian Knoll, Kerstin Hammernik, Erich Kobler, Thomas Pock, Michael~P Recht,
  and Daniel~K Sodickson.
\newblock Assessment of the generalization of learned image reconstruction and
  the potential for transfer learning.
\newblock \emph{Magnetic resonance in medicine}, 81\penalty0 (1):\penalty0
  116--128, 2019.

\bibitem[Liang et~al.(2021)Liang, Cao, Sun, Zhang, Van~Gool, and
  Timofte]{Swinir}
Jingyun Liang, Jiezhang Cao, Guolei Sun, Kai Zhang, Luc Van~Gool, and Radu
  Timofte.
\newblock Swinir: Image restoration using swin transformer.
\newblock In \emph{Proceedings of the IEEE/CVF international conference on
  computer vision}, pp.\  1833--1844, 2021.

\bibitem[Lin \& Heckel(2022)Lin and Heckel]{MRViT}
Kang Lin and Reinhard Heckel.
\newblock Vision transformers enable fast and robust accelerated mri.
\newblock In \emph{International Conference on medical imaging with deep
  learning}, pp.\  774--795. PMLR, 2022.

\bibitem[Lin \& Heckel(2023)Lin and Heckel]{DiverseTrain}
Kang Lin and Reinhard Heckel.
\newblock Robustness of deep learning for accelerated mri: benefits of diverse
  training data.
\newblock \emph{arXiv preprint arXiv:2312.10271}, 2023.

\bibitem[Lin et~al.(2025)Lin, Krainovic, Wang, and Heckel]{DataFilter}
Kang Lin, Anselm Krainovic, Kun Wang, and Reinhard Heckel.
\newblock Improving deep learning for accelerated mri with data filtering.
\newblock \emph{arXiv preprint arXiv:2508.13822}, 2025.

\bibitem[Liu et~al.(2021)Liu, Lin, Cao, Hu, Wei, Zhang, Lin, and Guo]{Swin}
Ze~Liu, Yutong Lin, Yue Cao, Han Hu, Yixuan Wei, Zheng Zhang, Stephen Lin, and
  Baining Guo.
\newblock Swin transformer: Hierarchical vision transformer using shifted
  windows.
\newblock In \emph{Proceedings of the IEEE/CVF international conference on
  computer vision}, pp.\  10012--10022, 2021.

\bibitem[Lu et~al.(2025)Lu, Jiang, Liu, Du, Jiang, Hong, Liu, He, Yuan, Wang,
  et~al.]{Moba}
Enzhe Lu, Zhejun Jiang, Jingyuan Liu, Yulun Du, Tao Jiang, Chao Hong, Shaowei
  Liu, Weiran He, Enming Yuan, Yuzhi Wang, et~al.
\newblock Moba: Mixture of block attention for long-context llms.
\newblock \emph{arXiv preprint arXiv:2502.13189}, 2025.

\bibitem[Lyu et~al.(2023)Lyu, Mei, Huang, Liu, Li, Yang, Liu, Dong, Dong, and
  Wu]{m4raw}
Mengye Lyu, Lifeng Mei, Shoujin Huang, Sixing Liu, Yi~Li, Kexin Yang, Yilong
  Liu, Yu~Dong, Linzheng Dong, and Ed~X Wu.
\newblock M4raw: A multi-contrast, multi-repetition, multi-channel mri k-space
  dataset for low-field mri research.
\newblock \emph{Scientific Data}, 10\penalty0 (1):\penalty0 264, 2023.

\bibitem[Peebles \& Xie(2023)Peebles and Xie]{DiT}
William Peebles and Saining Xie.
\newblock Scalable diffusion models with transformers.
\newblock In \emph{Proceedings of the IEEE/CVF international conference on
  computer vision}, pp.\  4195--4205, 2023.

\bibitem[Radford et~al.(2021)Radford, Kim, Hallacy, Ramesh, Goh, Agarwal,
  Sastry, Askell, Mishkin, Clark, et~al.]{CLIP}
Alec Radford, Jong~Wook Kim, Chris Hallacy, Aditya Ramesh, Gabriel Goh,
  Sandhini Agarwal, Girish Sastry, Amanda Askell, Pamela Mishkin, Jack Clark,
  et~al.
\newblock Learning transferable visual models from natural language
  supervision.
\newblock In \emph{International conference on machine learning}, pp.\
  8748--8763. PmLR, 2021.

\bibitem[Rao et~al.(2021)Rao, Zhao, Liu, Lu, Zhou, and Hsieh]{Dynamicvit}
Yongming Rao, Wenliang Zhao, Benlin Liu, Jiwen Lu, Jie Zhou, and Cho-Jui Hsieh.
\newblock Dynamicvit: Efficient vision transformers with dynamic token
  sparsification.
\newblock \emph{Advances in neural information processing systems},
  34:\penalty0 13937--13949, 2021.

\bibitem[Ronneberger et~al.(2015)Ronneberger, Fischer, and Brox]{Unet}
Olaf Ronneberger, Philipp Fischer, and Thomas Brox.
\newblock U-net: Convolutional networks for biomedical image segmentation.
\newblock In \emph{International Conference on Medical image computing and
  computer-assisted intervention}, pp.\  234--241. Springer, 2015.

\bibitem[Schlemper et~al.(2017)Schlemper, Caballero, Hajnal, Price, and
  Rueckert]{DeepCascade}
Jo~Schlemper, Jose Caballero, Joseph~V Hajnal, Anthony~N Price, and Daniel
  Rueckert.
\newblock A deep cascade of convolutional neural networks for dynamic mr image
  reconstruction.
\newblock \emph{IEEE transactions on Medical Imaging}, 37\penalty0
  (2):\penalty0 491--503, 2017.

\bibitem[Shazeer(2020)]{SwiGLU}
Noam Shazeer.
\newblock Glu variants improve transformer.
\newblock \emph{arXiv preprint arXiv:2002.05202}, 2020.

\bibitem[Shi et~al.(2016)Shi, Caballero, Husz{\'a}r, Totz, Aitken, Bishop,
  Rueckert, and Wang]{SubPixel}
Wenzhe Shi, Jose Caballero, Ferenc Husz{\'a}r, Johannes Totz, Andrew~P Aitken,
  Rob Bishop, Daniel Rueckert, and Zehan Wang.
\newblock Real-time single image and video super-resolution using an efficient
  sub-pixel convolutional neural network.
\newblock In \emph{Proceedings of the IEEE conference on computer vision and
  pattern recognition}, pp.\  1874--1883, 2016.

\bibitem[Siedler et~al.(2024)Siedler, Jakob, and Herold]{SCAMPI}
Thomas~M Siedler, Peter~M Jakob, and Volker Herold.
\newblock Enhancing quality and speed in database-free neural network
  reconstructions of undersampled mri with scampi.
\newblock \emph{Magnetic Resonance in Medicine}, 92\penalty0 (3):\penalty0
  1232--1247, 2024.

\bibitem[Souza et~al.(2018)Souza, Lucena, Garrafa, Gobbi, Saluzzi, Appenzeller,
  Rittner, Frayne, and Lotufo]{ccbrain}
Roberto Souza, Oeslle Lucena, Julia Garrafa, David Gobbi, Marina Saluzzi,
  Simone Appenzeller, Let{\'\i}cia Rittner, Richard Frayne, and Roberto Lotufo.
\newblock An open, multi-vendor, multi-field-strength brain mr dataset and
  analysis of publicly available skull stripping methods agreement.
\newblock \emph{NeuroImage}, 170:\penalty0 482--494, 2018.

\bibitem[Sriram et~al.(2020{\natexlab{a}})Sriram, Zbontar, Murrell, Defazio,
  Zitnick, Yakubova, Knoll, and Johnson]{E2EVarNet}
Anuroop Sriram, Jure Zbontar, Tullie Murrell, Aaron Defazio, C~Lawrence
  Zitnick, Nafissa Yakubova, Florian Knoll, and Patricia Johnson.
\newblock End-to-end variational networks for accelerated mri reconstruction.
\newblock In \emph{International conference on medical image computing and
  computer-assisted intervention}, pp.\  64--73. Springer, 2020{\natexlab{a}}.

\bibitem[Sriram et~al.(2020{\natexlab{b}})Sriram, Zbontar, Murrell, Zitnick,
  Defazio, and Sodickson]{GrappaNet}
Anuroop Sriram, Jure Zbontar, Tullie Murrell, C~Lawrence Zitnick, Aaron
  Defazio, and Daniel~K Sodickson.
\newblock Grappanet: Combining parallel imaging with deep learning for
  multi-coil mri reconstruction.
\newblock In \emph{Proceedings of the IEEE/CVF Conference on Computer Vision
  and Pattern Recognition}, pp.\  14315--14322, 2020{\natexlab{b}}.

\bibitem[Su et~al.(2024)Su, Ahmed, Lu, Pan, Bo, and Liu]{ROPE}
Jianlin Su, Murtadha Ahmed, Yu~Lu, Shengfeng Pan, Wen Bo, and Yunfeng Liu.
\newblock Roformer: Enhanced transformer with rotary position embedding.
\newblock \emph{Neurocomputing}, 568:\penalty0 127063, 2024.

\bibitem[Team et~al.(2024)Team, Riviere, Pathak, Sessa, Hardin, Bhupatiraju,
  Hussenot, Mesnard, Shahriari, Ram{\'e}, et~al.]{Gemma2}
Gemma Team, Morgane Riviere, Shreya Pathak, Pier~Giuseppe Sessa, Cassidy
  Hardin, Surya Bhupatiraju, L{\'e}onard Hussenot, Thomas Mesnard, Bobak
  Shahriari, Alexandre Ram{\'e}, et~al.
\newblock Gemma 2: Improving open language models at a practical size.
\newblock \emph{arXiv preprint arXiv:2408.00118}, 2024.

\bibitem[Uecker et~al.(2014)Uecker, Lai, Murphy, Virtue, Elad, Pauly,
  Vasanawala, and Lustig]{ESPIRiT}
Martin Uecker, Peng Lai, Mark~J Murphy, Patrick Virtue, Michael Elad, John~M
  Pauly, Shreyas~S Vasanawala, and Michael Lustig.
\newblock Espirit—an eigenvalue approach to autocalibrating parallel mri:
  where sense meets grappa.
\newblock \emph{Magnetic resonance in medicine}, 71\penalty0 (3):\penalty0
  990--1001, 2014.

\bibitem[Uecker et~al.(2015)Uecker, Ong, Tamir, Bahri, Virtue, Cheng, Zhang,
  and Lustig]{bart}
Martin Uecker, Frank Ong, Jonathan~I Tamir, Dara Bahri, Patrick Virtue,
  Joseph~Y Cheng, Tao Zhang, and Michael Lustig.
\newblock Berkeley advanced reconstruction toolbox.
\newblock In \emph{Proc. Intl. Soc. Mag. Reson. Med}, volume~23, pp.\ ~9, 2015.

\bibitem[Wang et~al.(2024)Wang, Guo, Wang, Cheng, Yu, and Wen]{PDAC}
Chong Wang, Lanqing Guo, Yufei Wang, Hao Cheng, Yi~Yu, and Bihan Wen.
\newblock Progressive divide-and-conquer via subsampling decomposition for
  accelerated mri.
\newblock In \emph{Proceedings of the IEEE/CVF Conference on Computer Vision
  and Pattern Recognition}, pp.\  25128--25137, 2024.

\bibitem[Wang et~al.(2020)Wang, Zhu, Green, Adam, Yuille, and Chen]{Axial}
Huiyu Wang, Yukun Zhu, Bradley Green, Hartwig Adam, Alan Yuille, and
  Liang-Chieh Chen.
\newblock Axial-deeplab: Stand-alone axial-attention for panoptic segmentation.
\newblock In \emph{European conference on computer vision}, pp.\  108--126.
  Springer, 2020.

\bibitem[Wang et~al.(2025)Wang, Guo, Chai, Zhou, Xu, and Jiang]{SDUM}
Puyang Wang, Pengfei Guo, Keyi Chai, Jinyuan Zhou, Daguang Xu, and Shanshan
  Jiang.
\newblock Sdum: A scalable deep unrolled model for universal mri
  reconstruction.
\newblock \emph{arXiv preprint arXiv:2512.17137}, 2025.

\bibitem[Wang et~al.(2022)Wang, Cun, Bao, Zhou, Liu, and Li]{Uformer}
Zhendong Wang, Xiaodong Cun, Jianmin Bao, Wengang Zhou, Jianzhuang Liu, and
  Houqiang Li.
\newblock Uformer: A general u-shaped transformer for image restoration.
\newblock In \emph{Proceedings of the IEEE/CVF conference on computer vision
  and pattern recognition}, pp.\  17683--17693, 2022.

\bibitem[Xin et~al.(2024)Xin, Ye, Axel, and Metaxas]{PromptMR+}
Bingyu Xin, Meng Ye, Leon Axel, and Dimitris~N Metaxas.
\newblock Rethinking deep unrolled model for accelerated mri reconstruction.
\newblock In \emph{European Conference on Computer Vision}, pp.\  164--181.
  Springer, 2024.

\bibitem[Yaman et~al.(2021)Yaman, Hosseini, and Ak{\c{c}}akaya]{ZS-SSL}
Burhaneddin Yaman, Seyed Amir~Hossein Hosseini, and Mehmet Ak{\c{c}}akaya.
\newblock Zero-shot self-supervised learning for mri reconstruction.
\newblock \emph{arXiv preprint arXiv:2102.07737}, 2021.

\bibitem[Yiasemis et~al.(2022)Yiasemis, Sonke, S{\'a}nchez, and
  Teuwen]{RecurrentVarNet}
George Yiasemis, Jan-Jakob Sonke, Clarisa S{\'a}nchez, and Jonas Teuwen.
\newblock Recurrent variational network: a deep learning inverse problem solver
  applied to the task of accelerated mri reconstruction.
\newblock In \emph{Proceedings of the IEEE/CVF conference on computer vision
  and pattern recognition}, pp.\  732--741, 2022.

\bibitem[Zamir et~al.(2022)Zamir, Arora, Khan, Hayat, Khan, and
  Yang]{Restormer}
Syed~Waqas Zamir, Aditya Arora, Salman Khan, Munawar Hayat, Fahad~Shahbaz Khan,
  and Ming-Hsuan Yang.
\newblock Restormer: Efficient transformer for high-resolution image
  restoration.
\newblock In \emph{Proceedings of the IEEE/CVF conference on computer vision
  and pattern recognition}, pp.\  5728--5739, 2022.

\bibitem[Zbontar et~al.(2018)Zbontar, Knoll, Sriram, Murrell, Huang, Muckley,
  Defazio, Stern, Johnson, Bruno, et~al.]{fastMRI}
Jure Zbontar, Florian Knoll, Anuroop Sriram, Tullie Murrell, Zhengnan Huang,
  Matthew~J Muckley, Aaron Defazio, Ruben Stern, Patricia Johnson, Mary Bruno,
  et~al.
\newblock fastmri: An open dataset and benchmarks for accelerated mri.
\newblock \emph{arXiv preprint arXiv:1811.08839}, 2018.

\bibitem[Zhang et~al.(2025)Zhang, Xiang, Huang, Wei, Xi, Zhu, and
  Chen]{Spargeattention}
Jintao Zhang, Chendong Xiang, Haofeng Huang, Jia Wei, Haocheng Xi, Jun Zhu, and
  Jianfei Chen.
\newblock Spargeattention: Accurate and training-free sparse attention
  accelerating any model inference.
\newblock \emph{arXiv preprint arXiv:2502.18137}, 2025.

\bibitem[Zhou et~al.(2025)Zhou, Pan, and Yang]{AST-v2}
Shihao Zhou, Jinshan Pan, and Jufeng Yang.
\newblock Learning an adaptive sparse transformer for efficient image
  restoration.
\newblock \emph{IEEE Transactions on Pattern Analysis and Machine
  Intelligence}, 2025.

\end{thebibliography}

\clearpage
\appendix
\raggedbottom
\section{Dataset Summary}

\begin{table}[H]
\caption{Summary of MRI datasets. 3D datasets are treated as multi-slice 2D data in selected anatomical views: CC-359 is converted to axial and sagittal views as two separate test sets, and AHEAD to the axial view. The AHEAD sequence contains five echoes, each forming a separate test set. Sources: FastMRI \citep{fastMRI}, Stanford 2D \citep{stanford2d}, CC-359 \citep{ccbrain}, M4Raw \citep{m4raw}, and AHEAD \citep{AHEAD}.}
\label{tab:dataset}
\centering
\fontsize{7.5pt}{9pt}\selectfont
\begin{tabular}{@{}llrrrrrrrr@{}}
\toprule
\textbf{Split} & \textbf{Dataset Name} & \textbf{Part} & \textbf{View} & \textbf{Sequence} & \textbf{Vendor} & \textbf{B0} & \textbf{Coils} & \textbf{Subj.} & \textbf{Slices} \\
\midrule
Train & FastMRI Brain & brain & axial & 2D FSE & Siemens & 3T/1.5T & 2–28 & 4469 & 70748 \\
\midrule
Test(ID) & FastMRI Brain & brain & axial & 2D FSE & Siemens & 3T/1.5T & 2–24 & 558 & 8852 \\
\midrule
\multirow{10}{*}[-4pt]{\parbox{1cm}{\textbf{Test (OOD)}}} & FastMRI Knee & knee & coronal & 2D FSE & Siemens & 3T/1.5T & 15 & 150 & 5427 \\
& Stanford 2D & mix & mix & 2D FSE & GE & 3T & 3–32 & 89 & 2037 \\
& CC-359(Axi) & brain & axial & 3D MP-RAGE & GE & 3T & 12 & 67 & 13802 \\
& CC-359(Sag) & brain & sagittal & 3D MP-RAGE & GE & 3T & 12 & 67 & 9490 \\
& M4Raw(GRE) & brain & axial & 2D GRE & Xingaoyi & 0.3T & 4 & 183 & 6588 \\
& AHEAD(Echo1) & brain & axial & 3D MP2RAGEME & Philips & 7T & 32 & 20 & 3240 \\
& AHEAD(Echo2) & brain & axial & 3D MP2RAGEME & Philips & 7T & 32 & 20 & 3240 \\
& AHEAD(Echo3) & brain & axial & 3D MP2RAGEME & Philips & 7T & 32 & 20 & 3240 \\
& AHEAD(Echo4) & brain & axial & 3D MP2RAGEME & Philips & 7T & 32 & 20 & 3240 \\
& AHEAD(Echo5) & brain & axial & 3D MP2RAGEME & Philips & 7T & 32 & 20 & 3240 \\
\bottomrule
\end{tabular}
\end{table}

\section{Additional Experimental Setups}
\label{sec:additional_setups}

\subsection{UEPS Architecture}
The unrolled architecture of UEPS consists of 4 cascades with a progressive resolution schedule of $64\times64$, $128\times128$, $256\times256$, and finally $320\times320$. The full-attention ViT baseline is built on DiT \citep{DiT} and incorporates 2D rotary position embeddings (RoPE) \citep{ROPE,Eva02} and SwiGLU feedforward networks \citep{SwiGLU}. The ViT backbone within each cascade is configured identically with a patch size of 8, 10 transformer layers, an embedding width of 512, 8 attention heads, and an MLP intermediate size of 1280, yielding approximately 120M parameters for the complete UEPS model. The sparse-attention ViT is implemented using FlexAttention \citep{FlexAttn}. Its 1D sparse attention is applied in cascades where the number of patches per image exceeds 256; otherwise, standard scaled dot-product attention is used (\eg, in the initial $64\times64$ low-resolution cascade). Within each 10-layer transformer, we interleave sparse- and full-attention layers: the first and last layers use full attention, while the remaining layers use sparse attention with $n=1$ adjacent row on each side.

\subsection{Training}
Our model was trained end-to-end on the FastMRI Brain \citep{fastMRI} training split with Mean Absolute Error (MAE) loss and the Adam optimizer for 30 epochs with a batch size of 8. The learning rate started with a linear warmup to the base value of $0.0003$ over the first 1\% of total iterations, and later decayed via a cosine annealing schedule down to 10\% of the base value. To accelerate training, we utilized Automatic Mixed Precision (AMP) alongside PyTorch's model compilation (\texttt{torch.compile}). To ensure fair evaluation, all models compared in our ablation and comparative experiments share the same training configuration, unless otherwise specified.

\subsection{Ablation Studies}
Using the ViT-based UE model as the baseline, we ablate progressive resolution and sparse attention by enabling each component independently and then jointly in UEPS. All four configurations use four unrolled cascades, approximately 120M parameters, and the same training protocol.

For the computational efficiency analysis in \cref{fig:speed}, we measure the inference runtime of all four variants at a fixed input resolution of $320\times320$. Patch sizes vary over 32, 16, 8, 4, 2, and 1, corresponding to 100, 400, 1600, 6400, 25600, and 102400 patches per full-resolution image, respectively.

\subsection{Multi-Resolution Comparison}
Training configuration was adjusted for all models in this study to accommodate the substantial memory and computational costs associated with Restormer \citep{Restormer} and ASTv2 \citep{AST-v2}. Specifically, all models were trained for 20 epochs with 8-coil training data, randomly sampled from the original multi-coil data (which contains a variable number of coils). Restormer \citep{Restormer} and ASTv2 \citep{AST-v2} have approximately 26M and 10M parameters, respectively, matching the configurations in their original publications.

\subsection{Sparse Attention Comparison}
To isolate the impact of the attention pattern itself, all models share the same parameter capacity of 120M. Furthermore, the sparsity patterns were configured to ensure that the effective receptive field (attention scope) for each patch is roughly comparable across models.

\section{Additional CSM Evaluation Results}

\Cref{tab:supp_csm_combination} provides the complete dataset-wise results corresponding to \cref{tab:rss_mvue}. Estimated CSMs are used to combine fully sampled coil images, and the combined images are evaluated against the RSS reference. PSNR is computed over foreground pixels; SSIM is obtained by averaging the full-image SSIM map within the same foreground mask. Slices with empty foreground masks are excluded.

\begin{table}[H]
\centering
\caption{Complete CSM evaluation results on the ID and ten OOD test sets. Values are mean $\pm$ standard deviation across valid slices. Average is computed over the ten OOD test sets. PSNR is reported in dB and SSIM in \%. Best values are bold and second best underlined.}
\label{tab:supp_csm_combination}
\small
\setlength{\tabcolsep}{4pt}
\renewcommand{\arraystretch}{1.08}
\resizebox{0.64\linewidth}{!}{%
\begin{tabular}{@{}llcccc@{}}
\toprule
 & & \multicolumn{4}{c}{CSM source} \\
\cmidrule(lr){3-6}
Dataset & Metric & BART & E2EVarNet & PDAC & NoVar \\
\midrule
\multirow{2}{*}{\parbox{1.6cm}{FastMRI\\Brain [ID]}} & PSNR$\uparrow$ & \textbf{44.95}\std{$\pm6.52$} & \underline{44.28}\std{$\pm4.59$} & 43.84\std{$\pm4.67$} & 41.98\std{$\pm4.33$} \\
 & SSIM$\uparrow$ & \textbf{99.00}\std{$\pm2.03$} & \underline{98.97}\std{$\pm2.30$} & 98.85\std{$\pm2.58$} & 98.85\std{$\pm2.48$} \\
\midrule
\multirow{2}{*}{\parbox{1.6cm}{CC-359\\Axi}} & PSNR$\uparrow$ & \textbf{41.86}\std{$\pm3.10$} & \underline{38.96}\std{$\pm3.30$} & 37.22\std{$\pm3.72$} & 38.83\std{$\pm2.94$} \\
 & SSIM$\uparrow$ & \textbf{98.74}\std{$\pm0.96$} & 98.34\std{$\pm1.02$} & 97.76\std{$\pm1.62$} & \underline{98.41}\std{$\pm0.92$} \\
\midrule
\multirow{2}{*}{\parbox{1.6cm}{CC-359\\Sag}} & PSNR$\uparrow$ & \textbf{42.02}\std{$\pm3.24$} & \underline{38.05}\std{$\pm3.02$} & 37.63\std{$\pm3.22$} & 36.89\std{$\pm2.52$} \\
 & SSIM$\uparrow$ & \textbf{98.82}\std{$\pm0.72$} & \underline{98.31}\std{$\pm0.79$} & 98.02\std{$\pm1.14$} & 98.21\std{$\pm0.75$} \\
\midrule
\multirow{2}{*}{\parbox{1.6cm}{M4Raw\\GRE}} & PSNR$\uparrow$ & \textbf{45.30}\std{$\pm1.61$} & \underline{44.40}\std{$\pm1.68$} & 44.38\std{$\pm1.77$} & 44.03\std{$\pm1.70$} \\
 & SSIM$\uparrow$ & \textbf{99.27}\std{$\pm0.25$} & \underline{99.20}\std{$\pm0.28$} & 99.18\std{$\pm0.32$} & 99.19\std{$\pm0.28$} \\
\midrule
\multirow{2}{*}{\parbox{1.6cm}{AHEAD\\Echo1}} & PSNR$\uparrow$ & \textbf{40.34}\std{$\pm2.23$} & \underline{37.09}\std{$\pm2.27$} & 35.61\std{$\pm2.22$} & 36.80\std{$\pm2.40$} \\
 & SSIM$\uparrow$ & \textbf{95.11}\std{$\pm2.14$} & 92.63\std{$\pm3.78$} & 89.82\std{$\pm6.38$} & \underline{92.89}\std{$\pm3.33$} \\
\midrule
\multirow{2}{*}{\parbox{1.6cm}{AHEAD\\Echo2}} & PSNR$\uparrow$ & \textbf{48.78}\std{$\pm4.03$} & \underline{44.40}\std{$\pm3.83$} & 43.91\std{$\pm3.85$} & 41.81\std{$\pm3.16$} \\
 & SSIM$\uparrow$ & \textbf{99.03}\std{$\pm1.04$} & \underline{98.34}\std{$\pm2.09$} & 97.75\std{$\pm3.10$} & 98.06\std{$\pm2.14$} \\
\midrule
\multirow{2}{*}{\parbox{1.6cm}{AHEAD\\Echo3}} & PSNR$\uparrow$ & \textbf{42.90}\std{$\pm3.71$} & \underline{36.19}\std{$\pm3.42$} & 33.89\std{$\pm2.90$} & 35.03\std{$\pm2.91$} \\
 & SSIM$\uparrow$ & \textbf{96.85}\std{$\pm3.15$} & \underline{93.33}\std{$\pm5.69$} & 89.81\std{$\pm8.52$} & 92.88\std{$\pm4.95$} \\
\midrule
\multirow{2}{*}{\parbox{1.6cm}{AHEAD\\Echo4}} & PSNR$\uparrow$ & \textbf{38.62}\std{$\pm4.04$} & \underline{33.74}\std{$\pm3.15$} & 32.07\std{$\pm3.17$} & 32.97\std{$\pm2.81$} \\
 & SSIM$\uparrow$ & \textbf{93.33}\std{$\pm5.43$} & \underline{88.37}\std{$\pm7.89$} & 83.14\std{$\pm11.68$} & 87.83\std{$\pm6.44$} \\
\midrule
\multirow{2}{*}{\parbox{1.6cm}{AHEAD\\Echo5}} & PSNR$\uparrow$ & \textbf{35.86}\std{$\pm4.08$} & \underline{32.35}\std{$\pm2.59$} & 30.98\std{$\pm2.91$} & 31.98\std{$\pm2.38$} \\
 & SSIM$\uparrow$ & \textbf{89.32}\std{$\pm7.25$} & \underline{84.09}\std{$\pm9.39$} & 77.93\std{$\pm13.50$} & 83.72\std{$\pm7.77$} \\
\midrule
\multirow{2}{*}{\parbox{1.6cm}{FastMRI\\Knee}} & PSNR$\uparrow$ & \textbf{38.76}\std{$\pm6.47$} & \underline{36.73}\std{$\pm6.80$} & 34.33\std{$\pm6.46$} & 34.85\std{$\pm5.40$} \\
 & SSIM$\uparrow$ & \textbf{96.55}\std{$\pm4.23$} & 95.11\std{$\pm6.28$} & 93.14\std{$\pm8.33$} & \underline{95.25}\std{$\pm5.79$} \\
\midrule
\multirow{2}{*}{\parbox{1.6cm}{Stanford\\2D}} & PSNR$\uparrow$ & \textbf{42.68}\std{$\pm6.25$} & \underline{39.67}\std{$\pm6.60$} & 38.22\std{$\pm6.17$} & 37.71\std{$\pm5.28$} \\
 & SSIM$\uparrow$ & \textbf{98.01}\std{$\pm2.60$} & \underline{96.70}\std{$\pm4.72$} & 95.81\std{$\pm6.41$} & 96.61\std{$\pm4.39$} \\
\midrule
\multirow{2}{*}{\parbox{1.6cm}{Average\\{}[OOD]}} & PSNR$\uparrow$ & \textbf{41.71} & \underline{38.16} & 36.82 & 37.09 \\
 & SSIM$\uparrow$ & \textbf{96.50} & \underline{94.44} & 92.24 & 94.31 \\
\bottomrule
\end{tabular}%
}
\end{table}

\section{Additional Zero-shot Transfer Benchmark Results}

\begin{figure}[H]
  \centering
  \includegraphics[width=\linewidth]{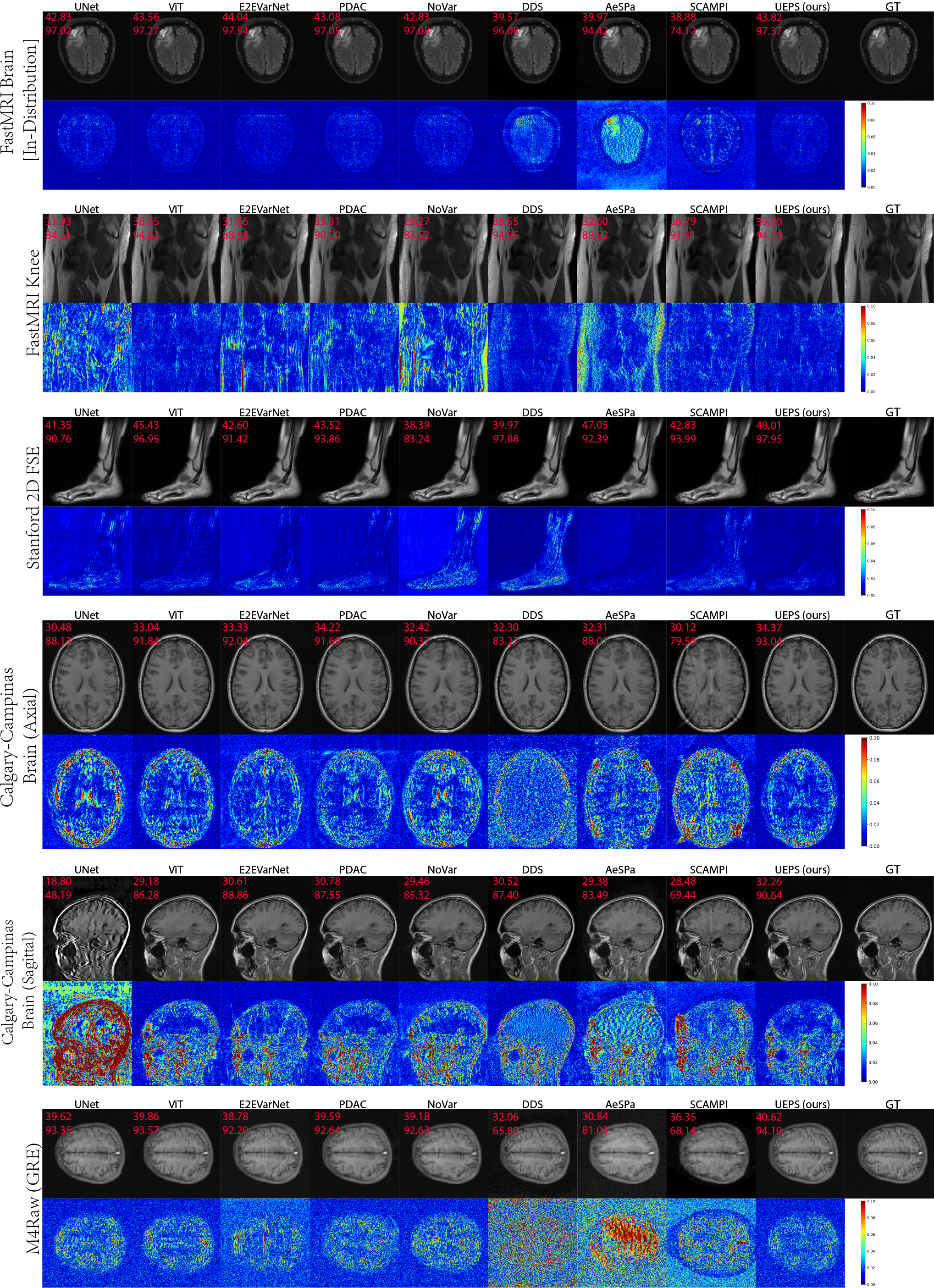}
  \caption{Additional reconstruction examples, each with PSNR/SSIM shown in the top-left corner (Part 1).}
  \label{fig:supp_vis1}
\end{figure}

\begin{figure}[H]
  \centering
  \includegraphics[width=\linewidth]{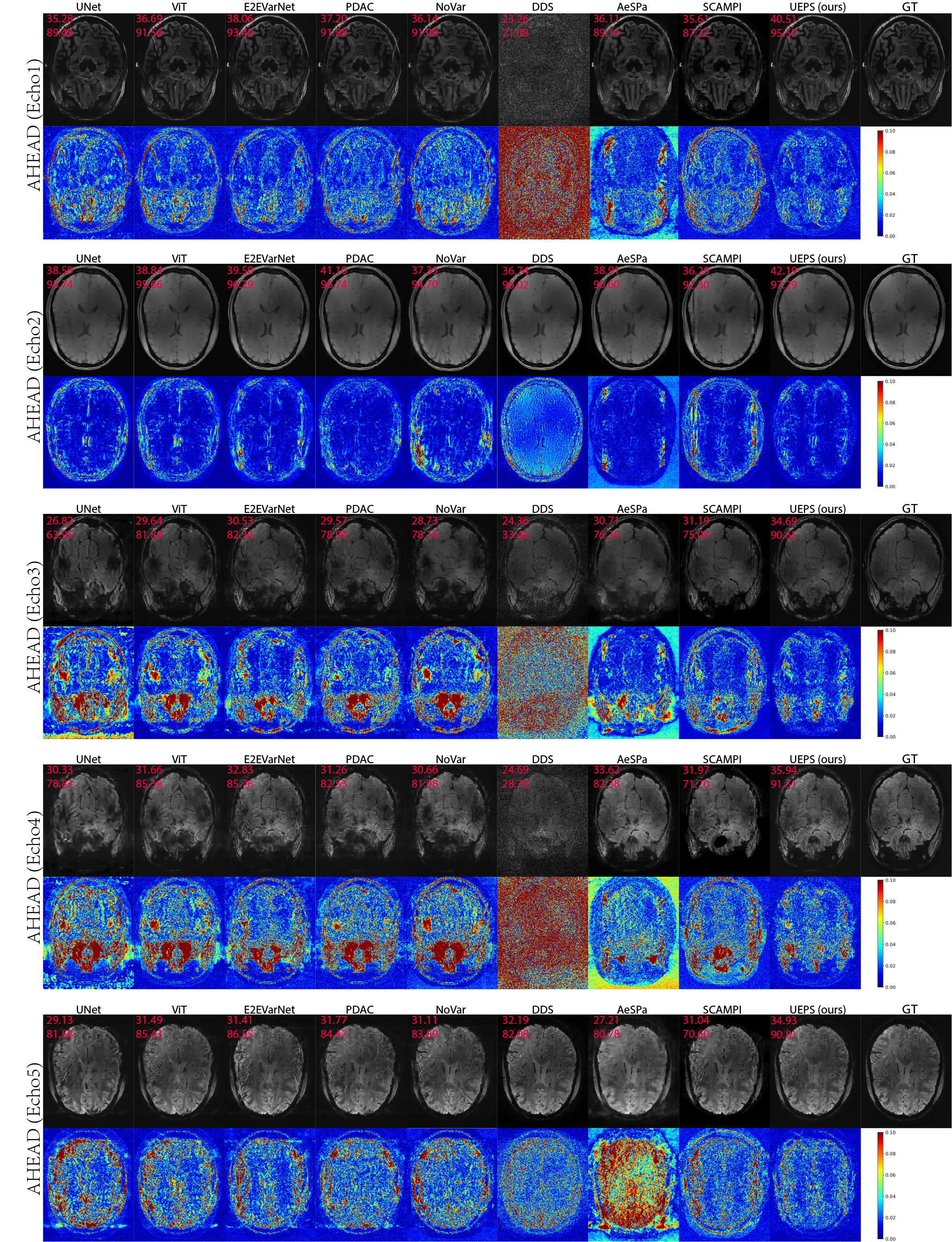}
  \caption{Additional reconstruction examples, each with PSNR/SSIM shown in the top-left corner (Part 2).}
  \label{fig:supp_vis2}
\end{figure}

Complete results for each test set, including quantitative results (\Cref{tab:supp_full_benchmark}) and visual examples (\Cref{fig:supp_vis1} and \cref{fig:supp_vis2}).

\begin{table}[H]
\caption{Complete results for the zero-shot transfer benchmark in \cref{fig:benchmark_psnr}. Average is computed over the ten OOD test sets. Best values are in bold and second best underlined.}
\label{tab:supp_full_benchmark}
\centering
\resizebox{\textwidth}{!}{%
\begin{tabular}{l l c c c c c c c c c}
\toprule
 & & \multicolumn{9}{c}{Model} \\
\cmidrule{3-11}
Dataset & Metric & UNet & ViT & E2EVarNet & PDAC & NoVar & DDS & AeSPa & SCAMPI & UEPS\suff{(Ours)} \\
\midrule
\multirow{2}{*}{\parbox{1.6cm}{FastMRI Brain [ID]}}
& PSNR(dB)$\uparrow$ & 40.58\std{$\pm3.02$} & 41.42\std{$\pm2.81$} & \textbf{41.93\std{$\pm2.86$}} & 41.03\std{$\pm2.81$} & 40.70\std{$\pm2.87$} & 35.91\std{$\pm5.37$} & 35.68\std{$\pm2.61$} & 36.48\std{$\pm2.42$} & \underline{41.87}\std{$\pm2.87$} \\
& SSIM(\%)$\uparrow$ & 95.61\std{$\pm3.49$} & 95.95\std{$\pm3.49$} & \textbf{96.27\std{$\pm3.34$}} & 95.69\std{$\pm3.63$} & 95.68\std{$\pm3.62$} & 85.52\std{$\pm21.49$} & 89.05\std{$\pm5.16$} & 76.48\std{$\pm10.54$} & \underline{96.15}\std{$\pm3.49$} \\
\midrule
\multirow{2}{*}{\parbox{1.6cm}{CC-359 Axi}}
& PSNR(dB)$\uparrow$ & 29.09\std{$\pm3.45$} & 30.99\std{$\pm3.52$} & 31.27\std{$\pm3.46$} & \underline{31.73}\std{$\pm3.48$} & 30.65\std{$\pm3.50$} & 29.17\std{$\pm5.42$} & 30.11\std{$\pm2.47$} & 29.83\std{$\pm2.52$} & \textbf{32.72\std{$\pm3.25$}} \\
& SSIM(\%)$\uparrow$ & 83.40\std{$\pm6.54$} & 87.34\std{$\pm5.19$} & \underline{87.86}\std{$\pm5.16$} & 87.36\std{$\pm5.09$} & 86.03\std{$\pm5.76$} & 74.10\std{$\pm25.72$} & 82.16\std{$\pm3.70$} & 73.17\std{$\pm7.98$} & \textbf{90.35\std{$\pm3.23$}} \\
\midrule
\multirow{2}{*}{\parbox{1.6cm}{CC-359 Sag}}
& PSNR(dB)$\uparrow$ & 27.36\std{$\pm5.08$} & 31.98\std{$\pm3.32$} & 32.86\std{$\pm3.32$} & \underline{32.99}\std{$\pm3.30$} & 31.57\std{$\pm3.47$} & 30.23\std{$\pm4.72$} & 31.82\std{$\pm2.13$} & 30.59\std{$\pm2.96$} & \textbf{34.45\std{$\pm3.33$}} \\
& SSIM(\%)$\uparrow$ & 77.01\std{$\pm13.56$} & 88.81\std{$\pm4.32$} & \underline{90.51}\std{$\pm4.05$} & 89.69\std{$\pm4.09$} & 87.65\std{$\pm5.24$} & 79.90\std{$\pm20.66$} & 83.43\std{$\pm3.14$} & 69.62\std{$\pm5.48$} & \textbf{92.01\std{$\pm3.03$}} \\
\midrule
\multirow{2}{*}{\parbox{1.6cm}{M4Raw GRE}}
& PSNR(dB)$\uparrow$ & 35.37\std{$\pm2.34$} & 36.37\std{$\pm2.08$} & 35.62\std{$\pm2.01$} & \underline{36.63}\std{$\pm1.93$} & 35.60\std{$\pm2.24$} & 26.73\std{$\pm9.07$} & 25.50\std{$\pm4.48$} & 32.64\std{$\pm1.50$} & \textbf{37.47\std{$\pm1.77$}} \\
& SSIM(\%)$\uparrow$ & 90.16\std{$\pm2.18$} & \underline{91.01}\std{$\pm1.85$} & 90.05\std{$\pm1.97$} & 90.31\std{$\pm1.93$} & 89.92\std{$\pm2.16$} & 58.03\std{$\pm29.52$} & 66.45\std{$\pm9.90$} & 60.72\std{$\pm3.89$} & \textbf{91.94\std{$\pm1.51$}} \\
\midrule
\multirow{2}{*}{\parbox{1.6cm}{AHEAD Echo1}}
& PSNR(dB)$\uparrow$ & 33.95\std{$\pm2.21$} & 35.67\std{$\pm2.24$} & \underline{35.94}\std{$\pm2.26$} & 35.80\std{$\pm2.23$} & 34.59\std{$\pm2.20$} & 24.50\std{$\pm3.56$} & 33.42\std{$\pm1.80$} & 35.05\std{$\pm1.53$} & \textbf{39.15\std{$\pm2.08$}} \\
& SSIM(\%)$\uparrow$ & 90.09\std{$\pm2.49$} & 92.40\std{$\pm1.95$} & \underline{93.18}\std{$\pm1.77$} & 92.05\std{$\pm2.18$} & 90.98\std{$\pm2.15$} & 36.19\std{$\pm21.01$} & 84.45\std{$\pm5.15$} & 82.94\std{$\pm5.43$} & \textbf{95.60\std{$\pm1.32$}} \\
\midrule
\multirow{2}{*}{\parbox{1.6cm}{AHEAD Echo2}}
& PSNR(dB)$\uparrow$ & 36.10\std{$\pm2.08$} & 37.53\std{$\pm2.03$} & 38.14\std{$\pm2.00$} & \underline{39.21}\std{$\pm2.10$} & 37.17\std{$\pm1.93$} & 31.08\std{$\pm5.31$} & 36.45\std{$\pm2.22$} & 36.38\std{$\pm1.68$} & \textbf{41.04\std{$\pm2.04$}} \\
& SSIM(\%)$\uparrow$ & 93.45\std{$\pm2.41$} & 94.67\std{$\pm1.91$} & 95.13\std{$\pm1.54$} & \underline{95.40}\std{$\pm1.95$} & 94.01\std{$\pm2.04$} & 68.71\std{$\pm30.43$} & 84.24\std{$\pm7.12$} & 90.39\std{$\pm2.30$} & \textbf{96.75\std{$\pm1.19$}} \\
\midrule
\multirow{2}{*}{\parbox{1.6cm}{AHEAD Echo3}}
& PSNR(dB)$\uparrow$ & 32.14\std{$\pm2.83$} & 34.03\std{$\pm2.72$} & \underline{34.33}\std{$\pm2.54$} & 33.86\std{$\pm2.68$} & 32.93\std{$\pm2.61$} & 26.98\std{$\pm4.25$} & 33.77\std{$\pm2.86$} & 32.82\std{$\pm2.04$} & \textbf{37.86\std{$\pm2.54$}} \\
& SSIM(\%)$\uparrow$ & 85.08\std{$\pm7.69$} & 89.13\std{$\pm4.15$} & \underline{89.63}\std{$\pm4.15$} & 87.85\std{$\pm4.62$} & 86.67\std{$\pm4.91$} & 50.04\std{$\pm27.62$} & 83.89\std{$\pm5.67$} & 77.22\std{$\pm4.68$} & \textbf{93.66\std{$\pm2.52$}} \\
\midrule
\multirow{2}{*}{\parbox{1.6cm}{AHEAD Echo4}}
& PSNR(dB)$\uparrow$ & 31.29\std{$\pm3.74$} & 33.51\std{$\pm3.51$} & \underline{34.03}\std{$\pm3.28$} & 33.48\std{$\pm3.37$} & 32.53\std{$\pm3.35$} & 26.44\std{$\pm3.84$} & 33.07\std{$\pm2.37$} & 32.86\std{$\pm1.57$} & \textbf{37.05\std{$\pm3.13$}} \\
& SSIM(\%)$\uparrow$ & 81.58\std{$\pm9.69$} & 87.17\std{$\pm5.23$} & \underline{88.23}\std{$\pm5.03$} & 85.90\std{$\pm5.69$} & 84.86\std{$\pm6.09$} & 45.90\std{$\pm25.30$} & 82.80\std{$\pm5.80$} & 70.48\std{$\pm5.53$} & \textbf{92.15\std{$\pm3.19$}} \\
\midrule
\multirow{2}{*}{\parbox{1.6cm}{AHEAD Echo5}}
& PSNR(dB)$\uparrow$ & 31.48\std{$\pm4.36$} & 33.86\std{$\pm3.93$} & 34.21\std{$\pm3.55$} & 33.62\std{$\pm3.75$} & 32.59\std{$\pm3.68$} & 25.45\std{$\pm3.30$} & 32.48\std{$\pm3.17$} & \underline{34.52}\std{$\pm2.23$} & \textbf{37.20\std{$\pm3.47$}} \\
& SSIM(\%)$\uparrow$ & 80.73\std{$\pm11.35$} & 87.10\std{$\pm5.71$} & \underline{88.11}\std{$\pm5.20$} & 85.80\std{$\pm6.14$} & 84.78\std{$\pm6.48$} & 37.31\std{$\pm19.64$} & 81.94\std{$\pm6.17$} & 72.16\std{$\pm4.93$} & \textbf{91.85\std{$\pm3.49$}} \\
\midrule
\multirow{2}{*}{\parbox{1.6cm}{FastMRI Knee}}
& PSNR(dB)$\uparrow$ & 35.98\std{$\pm5.31$} & \underline{39.89}\std{$\pm4.02$} & 38.40\std{$\pm4.69$} & 38.30\std{$\pm3.92$} & 36.67\std{$\pm4.63$} & 37.81\std{$\pm4.61$} & 33.71\std{$\pm4.12$} & 35.95\std{$\pm2.58$} & \textbf{40.07\std{$\pm3.93$}} \\
& SSIM(\%)$\uparrow$ & 86.74\std{$\pm9.46$} & \textbf{92.43\std{$\pm5.87$}} & 91.33\std{$\pm6.21$} & 90.38\std{$\pm6.59$} & 89.25\std{$\pm6.99$} & 89.19\std{$\pm12.75$} & 79.99\std{$\pm11.71$} & 76.91\std{$\pm14.43$} & \underline{92.38}\std{$\pm6.08$} \\
\midrule
\multirow{2}{*}{\parbox{1.6cm}{Stanford 2D}}
& PSNR(dB)$\uparrow$ & 36.31\std{$\pm6.00$} & \underline{38.84}\std{$\pm5.88$} & 38.54\std{$\pm5.44$} & 38.73\std{$\pm5.58$} & 37.81\std{$\pm4.94$} & 33.45\std{$\pm8.69$} & 33.23\std{$\pm7.34$} & 36.20\std{$\pm5.54$} & \textbf{40.52\std{$\pm5.85$}} \\
& SSIM(\%)$\uparrow$ & 90.34\std{$\pm8.70$} & \underline{93.11}\std{$\pm6.64$} & 92.96\std{$\pm6.98$} & 92.69\std{$\pm7.09$} & 91.87\std{$\pm6.30$} & 78.04\std{$\pm26.40$} & 76.55\std{$\pm14.50$} & 75.95\std{$\pm13.17$} & \textbf{94.39\std{$\pm6.45$}} \\
\midrule
\multirow{2}{*}{\parbox{1.6cm}{Average\\{}[OOD]}}
& PSNR(dB)$\uparrow$ & 32.91 & 35.27 & 35.33 & \underline{35.43} & 34.21 & 29.18 & 32.35 & 33.69 & \textbf{37.75} \\
& SSIM(\%)$\uparrow$ & 85.86 & 90.31 & \underline{90.70} & 89.74 & 88.60 & 61.74 & 80.59 & 74.95 & \textbf{93.11} \\
\bottomrule
\end{tabular}%
}
\end{table}

\end{document}